\documentclass[pra,floatfix,reprint,superscriptaddress]{revtex4-1}

\usepackage{graphicx}
\usepackage{units}
\usepackage{amsmath}
\usepackage{bbm}
\usepackage{braket}
\usepackage{xcolor}
\usepackage{import}
\usepackage{hyperref}
\usepackage[nameinlink]{cleveref}
\usepackage{mathtools}

\def\uarr{\mathord{\uparrow}}
\def\darr{\mathord{\downarrow}}
\def\defeq{\overset{\underset{\mathrm{def}}{}}{=}}
\def\beion{$^{9}\text{Be}^{+}\,\,$}
\def\mgion{$^{24}\text{Mg}^{+}\,\,$}

\def\modeOne{\text{$\nu_1$}}
\def\modeTwo{\text{$\nu_2$}}
\def\setOne{\mathcal{P}_{\text{inner}}}
\def\setTwo{\mathcal{P}_{\text{outer}}}
\def\eqnEndSpace{\,\,}
\def\rate{s^{-1}}

\newcommand{\KB}[2]{\Ket{\vphantom{#2}#1} 
	\hspace{-0.2em} \Bra{\vphantom{#1}#2}}

\hypersetup{
	colorlinks,
	linkcolor={blue!70!black},
	citecolor={blue!70!black},
	urlcolor={blue!70!black}
}

\begin{document}

\title{Quantum optimal control of the dissipative production of a maximally entangled state}

\author{Karl P. Horn}
\affiliation{Theoretische Physik, Universit\"{a}t Kassel, 
  Heinrich-Plett-Stra{\ss}e 40, D-34132 Kassel,
Germany}

\author{Florentin Reiter}
\affiliation{Department of Physics, Harvard University, 
  Cambridge, MA 02138, USA}

\author{Yiheng Lin}
\affiliation{CAS Key Laboratory of Microscale Magnetic Resonance and Department of Modern Physics, University of Science and Technology of China, Hefei 230026, China}
\affiliation{Synergetic Innovation Center of Quantum Information and Quantum Physics, University of Science and Technology of China, Hefei 230026, China}

\author{Dietrich  Leibfried}
\affiliation{National Institute of Standards and Technology, 
  Boulder, Colorado 80305, USA}

\author{Christiane P. Koch}
\affiliation{Theoretische Physik, Universit\"{a}t Kassel, 
  Heinrich-Plett-Stra{\ss}e 40, D-34132 Kassel,
Germany}
\email{christiane.koch@uni-kassel.de}

\date{\today}

\begin{abstract}
Entanglement generation can be robust against certain types of noise in approaches that deliberately incorporate dissipation into the system dynamics. The presence of additional dissipation channels may, however, limit 
fidelity and speed of the process.
Here we show how quantum optimal control techniques can be used to both speed up the entanglement generation and increase the fidelity in a realistic setup,
whilst respecting typical experimental
limitations. For the example of entangling two trapped ion qubits [Lin et al., Nature \textbf{504}, 415 (2013)], we find an improved fidelity by 
simply optimizing the polarization of the laser beams utilized in the
experiment. More significantly, an alternate combination of transitions between internal states of the ions, when combined with optimized polarization,
enables faster entanglement and  decreases
the error by an order of magnitude.
\end{abstract}

\maketitle

\section{Introduction}
Quantum devices aim to exploit 
the two essential elements of 
quantum physics, quantum coherence 
and entanglement, for practical applications. 
They require the implementation of 
a number of basic tasks such as state 
preparation or generation of entanglement, 
all the while preserving the relevant 
non-classical features at the level of 
device operation. 
The implementation of quantum tasks thus needs to be robust 
with respect to parameter fluctuations and 
external noise that is unavoidable in 
any real physical setup.

Loss of coherence and noise are commonly 
attributed to the coupling of the quantum 
system with its surrounding 
environment~\cite{breuer2002theory}. 
One strategy for realizing all necessary
tasks with sufficient accuracy is to 
perform the quantum operations at a 
time scale faster than the time scale 
at which the noise affects the system.
Quantum optimal control theory provides 
a set of tools to derive the corresponding 
protocols~\cite{glaser2015training} and  can be used to
identify the quantum speed limit~
\cite{caneva2009optimal,goerz2011quantum,patsch2018fast,sorensen2016exploring},
i.e., the shortest possible duration 
within which the operation can be 
carried out with a pre-specified fidelity. 

Nevertheless, there is a fundamental 
limit in that one cannot `beat' the noise,
particlularly, when its time scales are comparable to or
faster than the typical speed limits of 
the target operation.
An alternative is found in approaches that 
deliberately incorporate dissipation into 
the system dynamics, often referred to as 
quantum reservoir engineering~\cite{poyatos1996quantum}. 
The basic idea is to implement stochastic 
dynamics whose stationary state is non-classical. 
This is achieved by manipulating the coupling 
to the environment, or reservoir. In its simplest
form, a constant but switchable coupling is
realized by an electromagnetic field that drives
a transition to a state with fast decay~\cite{poyatos1996quantum}. 
The dynamics are described by
the quantum optical master equation~\cite{breuer2002theory}, 
and the system will 
eventually be driven into the fixed point of 
the corresponding Liouvillian 
~\cite{kraus2008preparation,verstraete2009quantum}. 

Applications of this basic idea  are many faceted
---its use has been suggested, for example in generating 
entanglement~
\cite{plenio1999cavity,benatti2003environment,vacanti2009cooling,wolf2011entangling,gonzalez2011entanglement,cho2011optical,kastoryano2011dissipative,bhaktavatsala2013dark,habibian2014stationary,fogarty2014quantum,bentley2014detection,morigi2015dissipative,aron2016photon,reiter2016scalable}, 
implementing universal quantum computing~\cite{verstraete2009quantum},
driving  phase transitions~\cite{cormick2012structural,diehl2008quantum,habibian2013bose}
and autonomous quantum error correction
~\cite{pastawski2011quantum,mirrahimi2014dynamically,reiter2017dissipative}.
Experimentally, the 
generation of non-classical states~\cite{kienzler2014quantum}, 
entangled states~\cite{krauter2011entanglement,lin2013dissipative,shankar2013autonomously,kimchi2016stabilizing}, 
and non-equilibrium quantum phases~
\cite{syassen2008strong,schindler2013quantum,barreiro2011open}
have successfully been demonstrated. 
Engineered dissipation can also be used 
towards a better understanding of open quantum system dynamics, by means of 
quantum simulation~\cite{barreiro2011open}. 

All of these examples testify to the fact that
dissipation can be a resource~\cite{verstraete2009quantum} for quantum technology. 
The ultimate performance bounds that can be 
reached with driven-dissipative dynamics under 
realistic conditions have, however, not yet 
been explored. 
While quantum reservoir engineering has been 
advocated for its robustness, its performance 
in a practical setting is compromised as soon 
as additional noise sources perturb the steady 
state or trap population flowing towards it.
 
This can be illustrated by examining the 
experiment described in Ref.~\cite{lin2013dissipative}.
For a \beion - \mgion - \mgion - \beion chain
occupying the same linear Paul trap,
the two \beion ions
were entangled via their collective motion using hyperfine 
electronic ground state levels as logical states.
Entanglement was achieved by applying a combination of  
laser and microwave transitions. 
This could be done in an either time-continuous 
manner or by repeating a fixed sequence of steps,
driving the system into a steady state, 
with the majority of population in the targeted, 
maximally entangled singlet state. 
Desired dissipation was brought into play 
by a combination of spin-motion coupling from a 
sideband laser, motion dissipation by sympathetically 
cooling cotrapped \mgion ions, and a repump laser 
which addresses the transition to a rapidly
decaying electronically excited state.
The sideband laser beams also lead to 
undesired pumping of spins, so-called
spontaneous emission. 
This resulted in population leakage and was 
the main source of error in that experiment~\cite{lin2013dissipative}. 

The simultaneous presence of both desired 
and undesired dissipation channels is rather generic. 
To harness the full power of dissipative 
entangled state preparation, one would like 
to exploit the former while mitigating the latter. 
Here, we use quantum optimal control theory
~\cite{glaser2015training} to address this problem. 
For the example of preparing two trapped ions in
a maximally entangled state~\cite{lin2013dissipative}, we 
ask whether entanglement can be generated faster
and more accurately when judiciously choosing a 
few key parameters. 
In order to keep in line with the experimental 
setup described in Ref.~\cite{lin2013dissipative},
we forego the usual assumption of time-dependent
pulses whose shapes are derived by quantum optimal 
control. 
Instead, we employ electromagnetic fields with 
constant amplitude and use tools from non-linear 
optimization to directly determine the best 
field strengths, detunings and polarizations. 
Our approach allows to not only determine the 
optimal values for these parameters, but also,
identify key factors that ultimately limit 
fidelity and speed of entanglement generation. 
Based on this insight, we explore an alternative
set of transitions and show that this scheme can 
outperform the original one both in terms of fidelity and speed. 

The paper is organized as follows. 
\Cref{sect:Model}
recalls the mechanism for entanglement generation in the experiment 
of Ref.~\cite{lin2013dissipative} and 
details the theoretical description of the corresponding
trapped ion system.
Optimization of the transitions used in Ref.~\cite{lin2013dissipative} is discussed in \cref{sect:Optim}. 
An alternative set of transitions is introduced in \cref{sect:twosid}, 
together with the optimization of the corresponding experimental parameters. 
We conclude in \Cref{sect:Conclusions}. 

\section{Model}
\label{sect:Model}

In this section we consider the system described in
Ref.~\cite{lin2013dissipative},
consisting of a linear Paul trap containing \beion ions and
\mgion ions, which interact mutually through their Coulomb
repulsion and with external electric fields.
A unitary idealization of these interactions
is summarized in the Hamiltonian $H$. The 
mechanism giving rise to dissipation in the state 
preparation process is spontaneous emission
after excitation of internal electronic
states of the ion by the external laser fields. 
The system dynamics is therefore described by 
the quantum optical master equation
in Lindblad form (with $\hbar = 1$),
\begin{equation}
	\partial_{t} \rho = \mathcal{L}\rho = -i \left[H,\rho\right] + \mathcal{L}_{\mathcal{D}} \rho\eqnEndSpace\,.
	\label{eq:QME}
\end{equation}
We refer to $\mathcal{L}_{\mathcal{D}}$ as
the (Lindblad) dissipator, which is
given by
\begin{align}
	\mathcal{L}_{\mathcal{D}} \rho = \sum_{k} \left( 
		L_{k} \rho L_{k}^{\dagger} - \frac{1}{2} \left[ 
		L_{k}^{\dagger} L_{k}, \rho \right] \right)\eqnEndSpace,
	\label{eq:LBpart}
\end{align}
where the sum over $k$ contains individual contributions 
due to sympathetic cooling, heating
and photon scattering
occurring during stimulated Raman processes and repumping
into an electronically excited state.

\begin{figure*}[tb]
	\includegraphics[width=\textwidth*9/10]{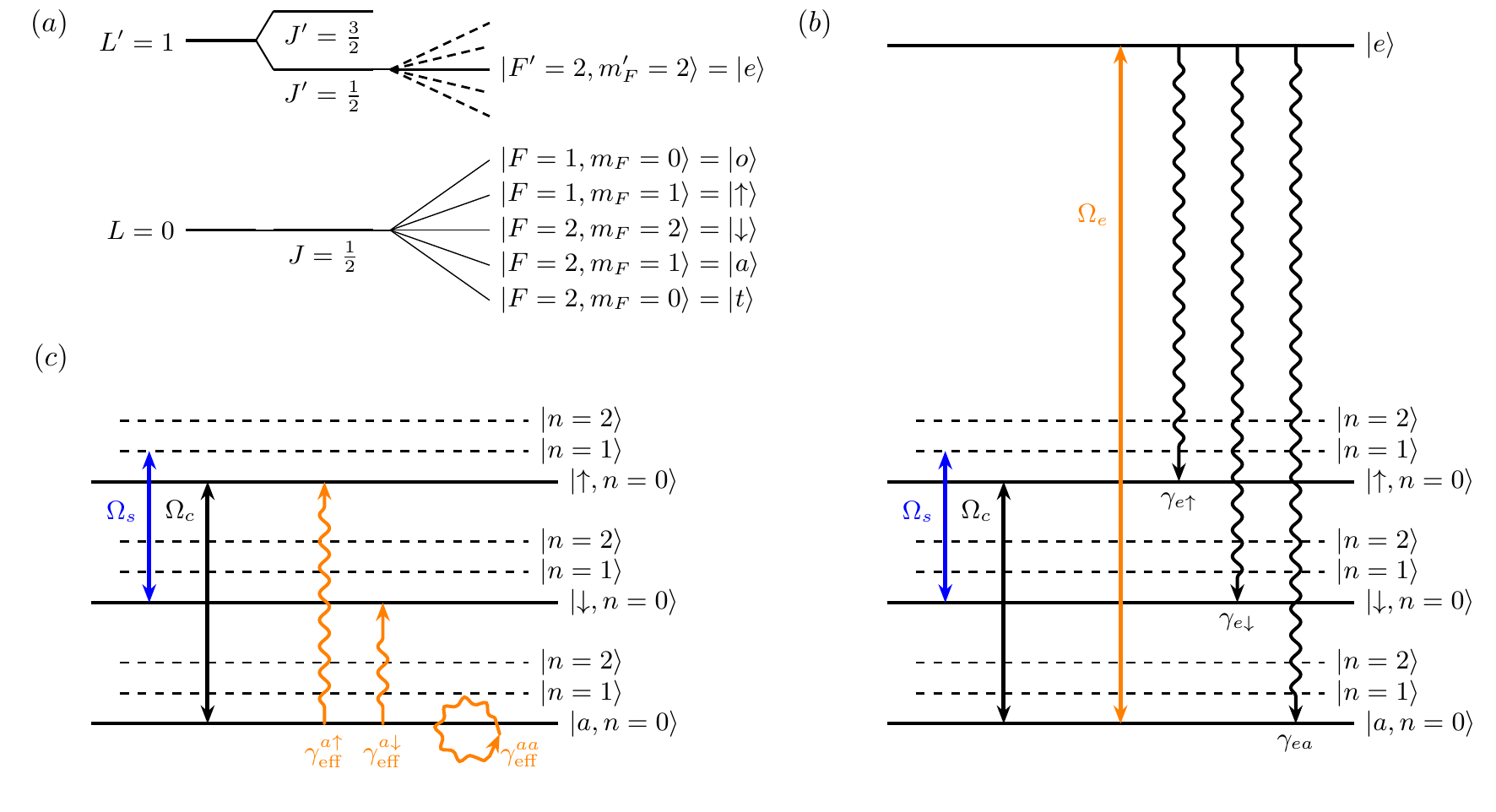}
	\caption{\label{fig:boulderTransitions} 
	(a) The fine and hyperfine structure of the
	electronic ground and first excited state of
	$^{9}\text{Be}^{+}$, including the most 
	relevant hyperfine levels. 
	(b) In the original scheme~\cite{lin2013dissipative}, a stimulated
	Raman blue sideband transition from 
	the $\Ket{\darr}$ to the $\Ket{\uarr}$ level ($\Omega_{s}$), 
	represented by a blue double-headed arrow, 
	a microwave carrier transition between the 
	$\Ket{a}$ and $\Ket{\uarr}$ levels ($\Omega_{c}$), represented by 
	a black double-headed arrow, and a repump 
	transition out of $\Ket{a}$ into the excited 
	level $\Ket{e}$ ($\Omega_{e}$), represented by an orange 
	double-headed arrow, are driven. 
	$\Ket{e}$ rapidly decays back into 
	$\Ket{a}$, $\Ket{\darr}$ and $\Ket{\uarr}$, 
	as represented by the black snaking lines.
	(c) In the picture after adiabatic elimination~\cite{lin2013dissipative},
	the excited level $\Ket{e}$ no longer explicitly 
	appears and effective decay, represented by 
	orange snaking lines, occurs directly out of $\Ket{a}$.
	In (b) and (c), $n$ refers to the occupation 
	number of the utilized vibrational mode.
}
\end{figure*}

\subsection{State space}
The model Hamiltonian $H$ accounts for the internal structure 
of two \beion ions as well as two vibrational modes of the trapped ion chain.
The state space of the considered system consists of the 
following tensor product structure,
\begin{align}
	(n_{qb1})\otimes (n_{qb2}) \otimes
	(n_{\modeOne}) \otimes (n_{\modeTwo})\eqnEndSpace.
	\label{eq:Structure}
\end{align}
In \cref{eq:Structure} $n_{qb1}$ and $n_{qb2}$ designate
hyperfine states of the \beion ions,
specified by the quantum numbers $F$ and their projections
$m_{F}$, obtained from coupling the total electronic angular
momentum quantum number $J$ with the nuclear spin quantum number 
$I$.
\Cref{fig:boulderTransitions}(a) highlights
the hyperfine states of interest, comprising of 
$\Ket{\darr} \defeq \Ket{S_{\nicefrac{1}{2}},F=2,m_{F}=2}$
and $\Ket{\uarr} \defeq \Ket{S_{\nicefrac{1}{2}},F=1,m_{F}=1}$,
the two hyperfine levels to entangle, as well as 
an auxiliary level
$\Ket{a} \defeq \Ket{S_{\nicefrac{1}{2}},F=2,m_{F}=1}$.
The neighbouring levels 
$\Ket{o}\defeq\Ket{S_{\nicefrac{1}{2}},F=1,m_{F}=0}$
and 
$\Ket{t}\defeq\Ket{S_{\nicefrac{1}{2}},F=2,m_{F}=0}$
are also accounted for in the model, since
these are predominantly populated by inadvertent scattering
processes.
In the following, the only electronically excited
state of interest will be $\Ket{e}\defeq\Ket{P_{\nicefrac{1}{2}},F'=2,m_{F}'=2}$.

$n_{\modeOne}$ and $n_{\modeTwo}$ are vibrational 
quantum numbers of two of the four shared 
motional modes of the trapped
ionic crystal along its linear axis.
Entanglement generation employs $\modeOne$, 
and sideband transitions utilizing 
this mode are essential for the
presented schemes.
Unless specifically required, the mode \modeTwo, which
is not utilized for entanglement
but is included in the model to account for off-resonant coupling, will
be suppressed notationally for the sake of simplicity.
It is assumed that the trap has an
axis of weakest confinement along which the 
four-ion string is aligned and that
the eight radial motional modes can be neglected,
since they are largely decoupled given the 
sideband laser configuration described in Ref.
~\cite{lin2013dissipative}.
\Cref{fig:boulderTransitions}(b)
shows three transitions that were driven on a single
\beion ion in Ref.~\cite{lin2013dissipative}. 
These belong to the
coherent part of \cref{eq:QME}, described by $H$, and one of them results in 
population of the electronically
excited state $\Ket{e}$ with subsequent dissipation
which is modeled by the incoherent part, $\mathcal{L_{D}}\rho$.
After adiabatic elimination, however, the transition to $\Ket{e}$
no longer appears in the
coherent part of \cref{eq:QME},
while the dissipative part
is modified by the result of the adiabatic
elimination to fully account for
the effective decay out of a electronic 
ground state hyperfine level instead~\cite{lin2013dissipative}. 
This is illustrated in  
\cref{fig:boulderTransitions}(c).

\subsection{Original scheme for entanglement preparation}
\label{ssect:originalScheme}

\begin{figure*}[tb]
	\includegraphics[scale=0.8]{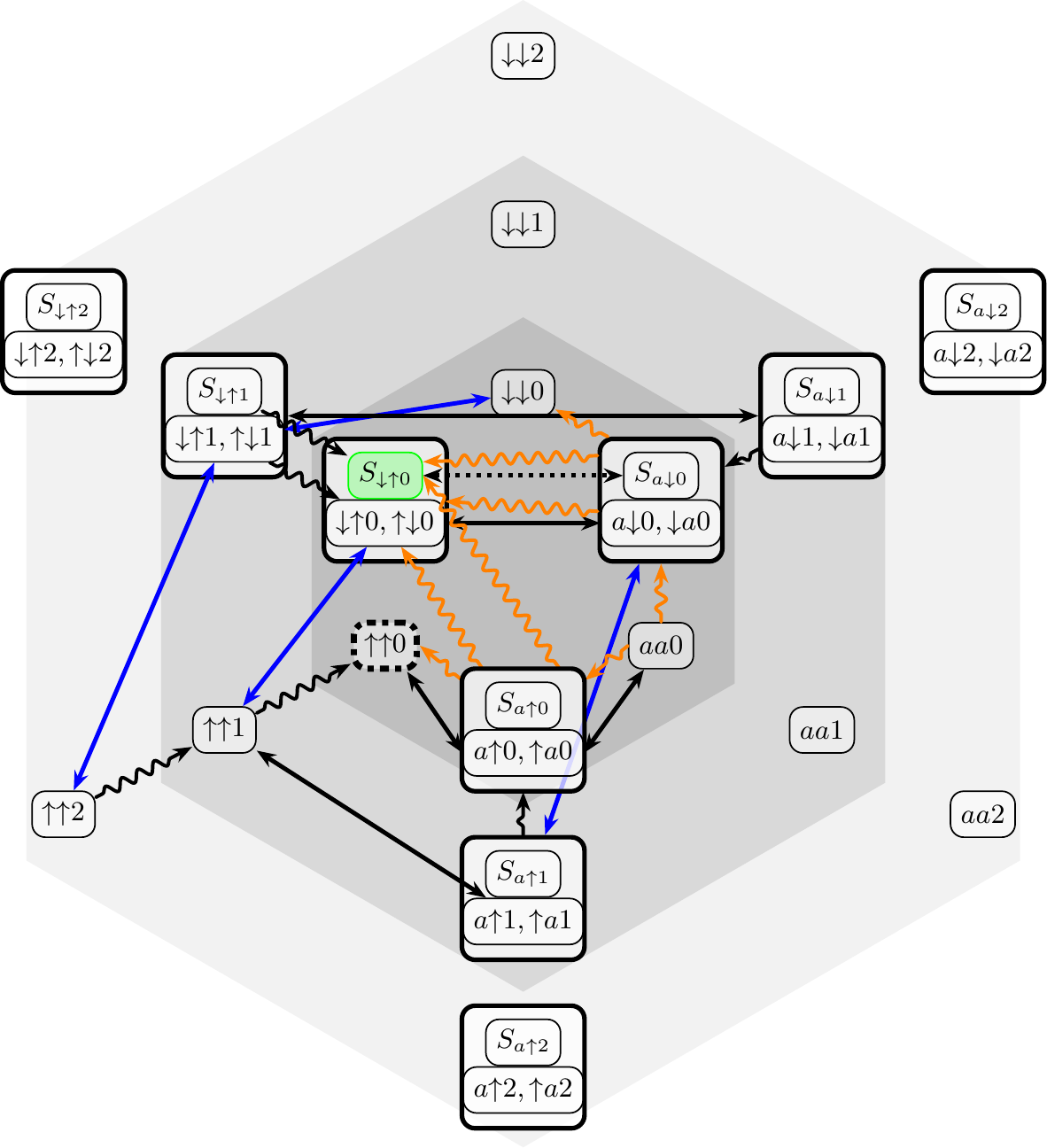}
	\caption{\label{fig:origMechanism} 
	Graphical overview of transitions needed for 
	steady state entanglement.
	For simplicity, only states within 
	the hyperfine subspace $\left\{ a,\darr,\uarr \right\}$ 
	and only the mode \modeOne~ are displayed.
	The vibrational quantum number of the 
	utilized mode increases radially outwards from the
	centre with shaded areas sharing the same quantum number.
	Carrier transitions between $\Ket{a}$ and 
	$\Ket{\uarr}$ at rate $\Omega_{\text{car},a,\uarr}$
	are represented by black double
	headed arrows.
	Sideband transitions between $\Ket{\darr}$ 
	and $\Ket{\uarr}$ at rate $\Omega_{\text{blue},\darr,\uarr}^{2p}$
	are represented by blue
	double headed arrows.
	Effective decay from $\Ket{a}$ 
	at the rates $\gamma_{a,f}^{\text{eff}}$
	for $f \in \left\{ a,\darr,\uarr \right\}$,
	is represented by orange snaking lines.
	Sympathetic cooling of the utilized 
	vibrational mode is represented by
	black snaking lines, whilst heating
	acts in the opposite direction and is
	not shown.
	For the sake of clarity,
	certain transitions are omitted
	and the 
	leaking between the hyperfine states
	is also not shown.
	}
\end{figure*}

As represented in \cref{fig:boulderTransitions}, 
the dissipative entanglement generation of Ref.~\cite{lin2013dissipative} uses three different types of fields to induce population flow in the state space.
The entanglement mechanism can be understood
by qualitatively tracing the flow of
population from state to state as indicated
in \cref{fig:origMechanism}.
Entangling the two \beion ions 
via their joint motion in the trap is made possible by utilizing
sideband transitions driven by Raman lasers. These change the internal
states of the \beion ions whilst simultaneously
exciting or de-exciting the utilized motional mode. In contrast, 
carrier transitions driven by a microwave field change the \beion
internal states only. Finally, a repump laser excites population to a short-lived electronically excited state. Specifically, in
Ref.~\cite{lin2013dissipative}, a single
sideband transition between 
$\Ket{\darr}$ and
$\Ket{\uarr}$,
a carrier transition
between 
$\Ket{a}$
and $\Ket{\uarr}$
and a repump transition between $\Ket{a}$ and
$\Ket{e}$ are used.
Figure \ref{fig:boulderTransitions} indicates
the transitions between the hyperfine levels
of interest for a single \beion ion. 
The above transitions can be driven simultaneously
and time-independently for the duration of the experiment or in a step-wise manner~\cite{lin2013dissipative}. Here, we focus on the continuous case, which resulted in a larger error.
Each \beion ion is affected by the driven transitions
independently and no individual addressing is required.
Starting with both \beion ions in an arbitrary state confined
to the hyperfine subspace $\left\{ a,\darr,\uarr \right\}$,
in the ideal case, this scheme always leads 
to a steady state in which the the population
is trapped in the singlet entangled state
between $\Ket{\darr}$ and $\Ket{\uarr}$, $\Ket{S_{\darr\uarr}} 
\defeq \frac{1}{\sqrt{2}} 
\left( \Ket{\darr\uarr} - \Ket{\uarr \darr} \right)$.
In the following, all singlet entangled states are
designated by $\Ket{S_{ij}} \defeq \frac{1}{\sqrt{2}}
\left( \Ket{i j} - \Ket{j i} \right)$,
whilst the triplet entangled states are designated
by $\Ket{T_{ij}} \defeq \frac{1}{\sqrt{2}}
\left( \Ket{i j} + \Ket{j i} \right)$, 
$\forall i,j \in \left\{ a,\darr,\uarr \right\}$.

Let us inspect in more detail the flow of
population from state to state 
in \cref{fig:origMechanism}.
Starting in $\Ket{\darr\darr n_{\modeOne}=0}$, for instance, 
it is possible to reach the target singlet entangled state
$\Ket{S_{\darr\uarr}}$ by two sideband transitions
leading to $\Ket{\uarr\uarr n_{\modeOne}}$, followed by a 
carrier transition into a combination of the
$\Ket{a\uarr n_{\modeOne}}$ and $\Ket{\uarr a n_{\modeOne}}$ states.
Population in the auxiliary state is driven by the repump laser into the electronically excited state from where it 
subsequently decays back into the electronic
ground state hyperfine subspace.
The process of electronic excitation and decay happens
sufficiently fast with respect to the other transitions,
that it can be regarded as `effective decay' directly
out of $\Ket{a}$, as depicted in 
\cref{fig:boulderTransitions}(b) and (c).
This decay drives the system into a combination
of $\Ket{\uarr \uarr n_{\modeOne}}$, the triplet entangled state
$\Ket{T_{\darr\uarr}}\otimes\Ket{n_{\modeOne}}$,
and the target state $\Ket{S_{\darr \uarr}}\otimes\Ket{n_{\modeOne}}$.
At any stage, sympathetic cooling can counteract the
excitations of the vibrational mode in the trap which are
caused by sideband transitions and heating.
Sympathetic cooling is induced by a different set of sideband lasers driving transitions only between internal states of the \mgion ions which share common motional modes with the \beion ions.
The carrier transition between $\Ket{a}$ and
$\Ket{\uarr}$ leads out of the target state
$\Ket{S_{\darr\uarr}}$ into 
$\Ket{S_{a\darr}}$.
This particular transition is highlighted 
specifically in \cref{fig:origMechanism} 
by a dotted black double headed arrow.

By ensuring that the two-photon Rabi frequency
$\Omega_{\text{blue},\darr,\uarr}^{2p}$
of the stimulated Raman sideband transition
between $\Ket{\darr}$ and $\Ket{\uarr}$ is
much larger than the carrier Rabi frequency
$\Omega_{\text{car},a,\uarr}$, the latter
transition can effectively be suppressed.
\Cref{fig:origMechanism} also highlights
the state $\Ket{\uarr\uarr 0}$ with a 
thick, dotted, black border, since 
the effective decay, proportional to
the square of the repump laser Rabi
frequency $\Omega_{\text{car},a,e}$
must be made sufficiently weak relative
to $\Omega_{\text{car},a,\uarr}$, in order
to prevent the trapping of population
in $\Ket{\uarr \uarr 0}$.
Consequently, a hierarchy of rates is
established in which the maximum
attainable two-photon Rabi frequency
of the stimulated Raman transition 
determines the maximal carrier
Rabi frequency between $\Ket{a}$ and 
$\Ket{\uarr}$, which in turn
determines the maximal repump Rabi
frequency between $\Ket{a}$ and $\Ket{e}$.

\subsection{Hamiltonian}
\label{ssect:Ham}
In the rotating wave approximation
and interaction picture,
the total system Hamiltonian is comprised of
the driven hyperfine transitions
\begin{align}
	H = \sum_{\text{type},i,f} H_{\text{type},i,f}
	\eqnEndSpace,
	\label{eq:Htot}
\end{align}
where the sum runs over specific triples
$\left(\text{type}, i,f \right)$, designating a
transition of type `red' or `blue' sideband 
or `carrier', between the initial 
and final hyperfine states $\Ket{i}$
and $\Ket{f}$.

Transitions of the carrier type 
between the ground state hyperfine levels
are driven by microwave fields
with a Hamiltonian of the form
\begin{align}
	H_{\text{car},i,f} = &\Omega_{\text{car},i,f} \big( \KB{f}{i}
    \otimes\mathbbm{1}_{\text{qb2}}\otimes\mathbbm{1}_{\modeOne}
    \otimes\mathbbm{1}_{\modeTwo}\nonumber \\ 
  &+ \mathbbm{1}_{\text{qb1}}\otimes\KB{f}{i}\otimes
    \mathbbm{1}_{\modeOne}\otimes
    \mathbbm{1}_{\modeTwo} \big)e^{-i\Delta_{\text{car},i,f}t} + \text{h.c.}\eqnEndSpace.
	\label{eq:Carriertrans}
\end{align}
Above, $\Omega_{\text{car},i,f}$ denotes the Rabi frequency and $\Delta_{\text{car},i,f}$ a small
detuning between the applied field and the transition energy
between $\Ket{i}$ and $\Ket{f}$.
Each identity operator $\mathbbm{1}_{j}$, with
$j \in \{\text{qb1, qb2, } \modeOne,\,\modeTwo \}$, is labelled
according to the subspace to which it corresponds.
A repump laser is required to drive 
transitions 
between ground and electronically excited 
hyperfine states.
These transitions therefore involve Hamiltonians of the form of \cref{eq:Carriertrans}, where
$\Ket{i}$ is a hyperfine ground state
level and $\Ket{f}= \Ket{e}$ is the addressed 
electronically excited hyperfine level.
Since population excited by this repumper decays very rapidly
into the hyperfine ground states, adiabatically eliminating
the excited state is well justified.

Ideally, a blue sideband transition between two hyperfine levels
$\Ket{i}$ and $\Ket{f}$, utilizing the motional mode
\modeOne, is represented  by 
\begin{align}
	H_{\text{blue},i,f} = &\Omega_{\text{blue},i,f} 
    \big( \KB{f}{i}\otimes \mathbbm{1}_{\text{qb2}}
    \otimes b^{+}\otimes\mathbbm{1}_{\modeTwo} \nonumber \\ 
  &+ \mathbbm{1}_{\text{qb1}}\otimes\KB{f}{i}
    \otimes b^{+}\otimes \mathbbm{1}_{\modeTwo} \big)e^{-i\Delta_{\text{blue},i,f}t} + \text{h.c.}\eqnEndSpace,
	\label{eq:SBtrans}
\end{align}
where $\Omega_{\text{blue},i,f}$ 
is the sideband Rabi frequency
and $\Delta_{\text{blue},i,f}$ 
a small detuning from the 
energy difference between $\Ket{i}$ and $\Ket{f}$
plus the energy of one quantum of \modeOne.
$b^{+}$ and $b$ denote the bosonic creation 
and annihilation operators which respectively excite and 
de-excite the harmonic mode \modeOne.
Analogously, the Hamiltonian of a red sideband 
transition takes the form of 
\cref{eq:SBtrans} but with the 
annihilation operator $b$
replacing the creation operators $b^\dagger$
and $\Delta_{\text{blue},i,f}$ replaced by 
$\Delta_{\text{red},i,f}$.

In the specific case of a stimulated Raman
sideband transition,
$\Omega_{\text{red/blue},i,f}$ in \cref{eq:SBtrans}
becomes $\Omega_{\text{red/blue},i,f}^{2p}$, a
two-photon Rabi frequency of a red/blue
sideband transition between $\Ket{i}$ and
$\Ket{f}$, given by
\begin{align}
	\Omega_{\text{red/blue},i,f}^{2p} = \eta_{\modeOne} \frac{\mu^{2} E_{r}E_{b}}{4 } 
	\sum_{k} \frac{\Bra{f} \bm{d}\cdot \bm{\varepsilon}_{r}\Ket{k}
	\Bra{k} \bm{d}\cdot \bm{\varepsilon}_{b}\Ket{i}}
			{\Delta_{k}\mu^{2}} \eqnEndSpace.
	\label{eq:SidebandRabifreq}
\end{align}
In the following we assume Lamb-Dicke parameters of
$\eta_{\modeOne}=0.180$ and $\eta_{\modeTwo}=0.155$
for the utilized ($\modeOne$) and off-resonant 
($\modeTwo$) motional modes, respectively \cite{lin2013dissipative}. 
Above, $E_{r}$ and $E_{b}$ are the field strengths of the 
lower (red) and higher (blue) frequency Raman laser beams 
which have 
polarizations $\bm{\varepsilon}_{r}$ and
$\bm{\varepsilon}_{b}$, expressed in spherical components as
$\bm{\varepsilon}_{r}=(r_{-},r_{0},r_{+})$
and $\bm{\varepsilon}_{b}=(b_{-},b_{0},b_{+})$,
respectively.
$\bm{d}$ is the dipole operator for the \beion ions
(also expressed in the spherical basis)
and the sum runs over all hyperfine levels $\Ket{k}$
in the electronically excited states $P_{\nicefrac{1}{2}}$
and $P_{\nicefrac{3}{2}}$.
The laser frequencies are shifted, such that
the ground state to excited state transitions
are detuned by $\Delta_{e}$
and $\Delta_{e} + f_{P}$ below the
$S_{\nicefrac{1}{2}} \leftrightarrow P_{\nicefrac{1}{2}}$
and $S_{\nicefrac{1}{2}}\leftrightarrow P_{\nicefrac{3}{2}}$ 
resonances, respectively.
$f_{P} \approx \unit[197.2]{GHz}$ is the fine structure splitting between 
$P_{\nicefrac{1}{2}}$ and $P_{\nicefrac{3}{2}}$.
For the detuning between $\Ket{i}$ and an individual excited state
hyperfine level $\Ket{k}$, the hyperfine splitting 
is neglected such that
\[
	\Delta_{k} \approx 
	\begin{cases}
		\Delta_{e}, & 
		\text{ if } \Ket{k} \in P_{\nicefrac{1}{2}} \\
		\Delta_{e}+f_{P}, & \text{ if } \Ket{k} \in  P_{\nicefrac{3}{2}}\,\,.
	\end{cases}
\]
\Cref{eq:SidebandRabifreq} utilizes
a characteristic stretched state
transition matrix element,
$\mu \defeq \Bra{P_{\nicefrac{3}{2}}, F=3, m_{F}=3} 
d_{+} \Ket{S_{\nicefrac{1}{2}}, F=2,m_{F}=2}$ 
to properly scale 
a given reduced matrix element, 
$\Bra{f}\bm{d}\cdot\varepsilon\Ket{i}/\mu$,
with $d_{+}$, the right circular
component of the dipole operator.
The Wigner-Eckart theorem
\cite{sobel'man1967introduction}
and Breit-Rabi formula 
\cite{woodgate1970elementary}
can then be used to
express an arbitrary transition matrix element
$\Bra{f}\bm{d}\cdot\bm{\varepsilon}\Ket{i}$
between two hyperfine levels $\Ket{i}$ and $\Ket{f}$. 

To accurately model the system dynamics, 
it is necessary to account for the undesired 
off-resonant coupling of a given sideband transition
described by \cref{eq:SBtrans} to an additional mode
\modeTwo, given by
\begin{align}
	H_{\text{blue},i,f}^{\modeTwo} = 
	& \frac{\eta_{\modeTwo}}{\eta_{\modeOne}} \Omega_{\text{blue},i,f}^{2p} 
  \big( \KB{f}{i}\otimes \mathbbm{1}_{\text{qb2}}
    \otimes \mathbbm{1}_{\modeOne} \otimes c^{+} \nonumber \\ 
  &+ \mathbbm{1}_{\text{qb1}} \otimes\KB{f}{i}\otimes 
    \mathbbm{1}_{\modeOne} \otimes c^{+} \big) \nonumber \\
    & \quad \quad \times e^{-i(\delta -\Delta_{\text{blue},i,f)}t} + \text{h.c.}\eqnEndSpace,
	\label{eq:SBORtrans}
\end{align}
in the case of a blue sideband transition.
In \cref{eq:SBORtrans}, $\delta$ is the detuning
between the utilized mode $\modeOne$ and 
$\modeTwo$, which couples off-resonantly.
In the case of a red sideband transition,
the off-resonant coupling takes the form
of \cref{eq:SBORtrans} under interchange
of the annihilation and creation operators
of the harmonic oscillator describing
the \modeTwo~motional mode, 
$c$ and $c^{+}$, and replacement of
$\Delta_{\text{blue},i,f}$ by 
$\Delta_{\text{red},i,f}$, respectively.

\subsection{Lindblad operators}

Incoherent processes taking place alongside
the driven transitions appear in the dissipative
part $\mathcal{L_{D}}$ in \cref{eq:QME}, which is comprised
of individual contributions modelled by the Lindblad (jump) 
operators $L_{k}$ in \cref{eq:LBpart}. An effective operator formalism~\cite{gardiner2004quantum} allows to adiabatically
eliminate the hyperfine excited state addressed by the
repump laser.
It leads to Lindblad operators of the form~\cite{reiter2012effective}
\begin{align}
	L_{\text{rep},i,f}^{(1)} &= \sqrt{\gamma_{if}^{\text{eff}}} \KB{f}{i} \otimes 
    \mathbbm{1}_{\text{qb2}} \otimes \mathbbm{1}_{\modeOne} 
    \otimes \mathbbm{1}_{\modeTwo}
	\label{eq:GammaRepFirst}
	\\
	L_{\text{rep},i,f}^{(2)} &= \sqrt{\gamma_{if}^{\text{eff}}} 
    \mathbbm{1}_{\text{qb1}} \otimes \KB{f}{i} 
    \otimes \mathbbm{1}_{\modeOne} \otimes \mathbbm{1}_{\modeTwo} \eqnEndSpace,
	\label{eq:GammaRepSecond}
\end{align}
with effective rates 
\begin{align}
	\gamma_{if}^{\text{eff}} = \gamma_{ef} \frac{4\Omega_{\text{car},i,e}^{2}}{\gamma^{2}}\eqnEndSpace,
	\label{eq:EffectiveRates}
\end{align}
where $\Ket{e}$ is the intermediate, rapidly decaying, 
electronically excited state,
$\Omega_{\text{car},i,e}$ the repump 
Rabi
frequency,
$\gamma_{ef}$ the decay rate from $\Ket{e}$ into the
hyperfine ground state $\Ket{f}$ and
$\gamma = \sum_{f'} \gamma_{ef'}$
the total decay rate out of $\Ket{e}$ into a subspace
of hyperfine ground states.

Similarly to \cref{eq:GammaRepFirst,eq:GammaRepSecond},
leaking between ground-state hyperfine levels
due to stimulated Raman sideband transition
acts on both beryllium ions according to
\begin{align}
	L_{\text{sid},i,f}^{(1)} &= \sqrt{\Gamma_{if}} \KB{f}{i} \otimes 
    \mathbbm{1}_{\text{qb2}} \otimes 
    \mathbbm{1}_{\modeOne} \otimes \mathbbm{1}_{\modeTwo} \\ 
  L_{\text{sid},i,f}^{(2)} &= \sqrt{\Gamma_{if}} \mathbbm{1}_{\text{qb1}}
    \otimes \KB{f}{i} \otimes \mathbbm{1}_{\modeOne}
    \otimes \mathbbm{1}_{\modeTwo}\,. 
	\label{eq:StimRamanPhotonScattering}
\end{align}
The scattering rate $\Gamma_{if}$
between an initial hyperfine ground state 
$\Ket{i}$ and a final hyperfine ground state
$\Ket{f}$, due to a single laser beam is
given by the Kramers-Heisenberg formula
\begin{align}
	\Gamma_{i f} = \Gamma_{i \rightarrow f} = \frac{\left|E\right|^{2}\mu^{2}}{4} \gamma \left| \sum_{k} 
		\frac{a_{i f}^{(k)}}{\Delta_{k}} \right|^{2}\eqnEndSpace,
	\label{eq:Kramers}
\end{align}
where
\begin{align}
	a_{i f}^{(k)} = a_{i \rightarrow f}^{(k)} = \sum_{q\in \left\{+,0,-\right\}} \frac{\Bra{f} d_{q} \Ket{k}}{\mu}
		\frac{\Bra{k} \bm{d} \cdot \bm{\varepsilon} \Ket{i}}{\mu}
	\label{eq:aif}
\end{align}
is the two-photon transition amplitude between $\Ket{i}$
and $\Ket{f}$.
As in \cref{eq:SidebandRabifreq}, $k$ runs over
all states $\Ket{k}$ belonging to 
the \beion ion $P_{\nicefrac{1}{2}}$ and
$P_{\nicefrac{3}{2}}$ manifolds.
Again, it sufficies to approximate
the $\Delta_{k}$ of 
$k \in P_{\nicefrac{1}{2}}, P_{\nicefrac{3}{2}}$
as $\Delta_{e}$
and $\Delta_{e}+f_{P}$, respectively. 
Rayleigh scattering is modelled by a Pauli $\sigma_{z}$
matrix between pairs of levels. 
Acting at the rate
\begin{align}
	\phi_{i f} = \frac{\left|E\right|^{2}\mu^{2}}{4} \gamma \left| \sum_{k} 
    \left( \frac{a_{i i}^{(k)}}{\Delta_{k}} 
    -\frac{a_{f f}^{(k)}}{\Delta_{k}} \right) \right|^{2}\eqnEndSpace,
	\label{eq:RayleighKramers}
\end{align}
Rayleigh scattering is only of concern between
the $\Ket{\darr}$ and $\Ket{\uarr}$ levels
and in most cases negligibly small.

Sympathetic cooling is achieved by using stimulated Raman
laser cooling and 
can be made to affect either or
both of the considered 
motional modes according to 
\begin{align}
  L_{\text{cool},\modeOne} &= \sqrt{\kappa_{c,\modeOne}} 
    \mathbbm{1}_{\text{qb1}}\otimes\mathbbm{1}_{\text{qb2}}
    \otimes b\otimes\mathbbm{1}_{\modeTwo} \\
	L_{\text{cool},\modeTwo} &= \sqrt{\kappa_{c,\modeTwo}} 
    \mathbbm{1}_{\text{qb1}}\otimes\mathbbm{1}_{\text{qb2}}
    \otimes\mathbbm{1}_{\modeOne}\otimes c
	\label{eq:SympCool}\eqnEndSpace,
\end{align}
where the cooling rates $\kappa_{c,\modeOne}$ 
and $\kappa_{c,\modeTwo}$
are governed by the field strengths of the repump and
stimulated Raman lasers acting on the magnesium ions.

Heating acts on all motional modes. 
It is caused by spontaneous emission occuring during the magnesium 
sideband Raman transitions, as well as photon recoil
from spontaneous emission and also the anomalous heating
of the ion trap.
The total heating can be modelled by
\begin{align}
  L_{\text{heat},\modeOne} &= \sqrt{\kappa_{h,\modeOne}} 
    \mathbbm{1}_{\text{qb1}}\otimes\mathbbm{1}_{\text{qb2}}\otimes 
    b^{\dagger}\otimes\mathbbm{1}_{\modeTwo} \\
  L_{\text{heat},\modeTwo} &= \sqrt{\kappa_{h,\modeTwo}} 
    \mathbbm{1}_{\text{qb1}}\otimes\mathbbm{1}_{\text{qb2}}\otimes 
    \mathbbm{1}_{\modeOne} \otimes c^{\dagger}\eqnEndSpace,
	\label{eq:Heating}
\end{align}
for a set of given heating rates $\kappa_{h,\modeOne}$
and $\kappa_{h,\modeTwo}$.

\section{Optimizing the original scheme}
\label{sect:Optim}

The goal of optimization 
is to maximize 
the 
population in the target state 
$\Ket{S_{\darr\uarr}}$.
To this end, the final time $T$ is defined as the
time at which the peak population in the
target state is reached and 
all driving fields 
can be turned off.
The target state population at final time is 
defined as the fidelity $F$
and correspondingly the error 
as
$\epsilon \defeq 1 - F$.
The peak population at the final time is an appropriate
quantity to observe, since the stability
of the ionic hyperfine ground states causes
the system to remain in its entangled state
for a long time after all driving fields 
have been turned off.

In the following, the system degrees of freedom 
available for control are introduced and categorised
into two collections
in preparation for the optimization scheme discussed
below.
In contrast to a straightforward parameter optimization
of all degrees of freedom, the specialised
optimization scheme presented here is
less susceptible to running into local minima 
and demonstrates reliable and 
fast convergence.

\subsection{Optimization parameters}
\label{ssect:OptPar}

As previously discussed, the limitations
of the original scheme~\cite{lin2013dissipative} are fundamentally
linked to the physical process of the stimulated Raman
sideband transition.
The two-photon Rabi frequency 
$\Omega_{\text{blue},\darr,\uarr}^{2p}$
associated with this transition
should be made as large as possible to drive the
system towards the desired target state whilst ensuring
that the unfavourable transition between $\Ket{S_{\darr\uarr}}$
and $\Ket{S_{a\darr}}$ is suppressed.
Consequently, the carrier transition Rabi
rate $\Omega_{\text{car},a,\uarr}$ and in
turn the repump transition Rabi rate 
governing the effective decay out of $\Ket{a}$ 
are limited, bottlenecking the flow
of population into the target state.

\Cref{eq:SidebandRabifreq,eq:Kramers,eq:aif}
show that merely increasing
the field strengths of the sideband lasers
has the adverse side effect of also increasing
the chance of photon scattering and therefore
the rates of leaking between hyperfine 
ground states.
As such, a safe way of increasing the
field strength of the sideband lasers
is to compensate by increasing the
detuning $\Delta_{e}$
from the excited state manifold,
since the two-photon Rabi 
frequency scales inversely with the detuning whilst
the scattering rates between hyperfine
states scale with the square of the inverse detuning.
The field strengths required 
to significantly increase the 
two-photon Rabi frequency
whilst minimising the associated 
scattering rates
are, however, beyond current experimental capabilities
\cite{ozeri2007errors}.
A third option is given by the polarization
of the two stimulated Raman sideband laser beams
$\bm{\varepsilon}_{r}$ and $\bm{\varepsilon}_{b}$,
which have a great impact on both $\Omega_{\text{blue},\darr,\uarr}^{2p}$ and also $\left\{ \Gamma_{if} \right\}$.

The tunable parameters $E_{r}$ and $E_{b}$,
$\bm{\varepsilon}_{r}$ and $\bm{\varepsilon}_{b}$
and $\Delta_{e}$,
appearing in
\Cref{eq:SidebandRabifreq,eq:Kramers,eq:aif}
constitute a first set of parameters defined as 
\begin{align}
	\setOne \defeq \left\{ E_{r},E_{b},\bm{\varepsilon}_{r},
\bm{\varepsilon}_{b}, \Delta_{e} \right\}
	\label{eq:setOne}\eqnEndSpace.
\end{align}
These are directly associated with the stimulated 
Raman sideband
transition.
The two-photon Rabi frequency 
$\Omega_{\text{blue},\darr,\uarr}^{2p}$
scales with the product of field strengths
$E_{r}E_{b}$, whilst the scattering 
rates due to each laser beam
scale with $\left|E_{r/b}\right|^{2}$, the magnitude of the field
strength squared.
The polarization is split into its three 
spherical components, 
$\bm{\varepsilon} = (\varepsilon_{-}, \varepsilon_{0}, \varepsilon_{+})$ 
where $\varepsilon_{i} \in \left[ -1,1 \right],\eqnEndSpace
\forall i \in \left\{ -,0,+ \right\}$ and
with
\begin{align}
	\left| \varepsilon_{-} \right|^{2} + 
	\left| \varepsilon_{0} \right|^{2} + 
	\left| \varepsilon_{+} \right|^{2} = 1\eqnEndSpace.
	\label{eq:polarization}\
\end{align}
Due to the normalisation of the spherical
components, each polarization
possesses two degrees of freedom
which can be represented as
the azimuthal and polar angles on the
unit sphere.

A given configuration
of $\setOne$ fully determines the resulting
two-photon Rabi frequency
$\Omega_{\text{blue},\darr,\uarr}^{2p}$
and all leakage rates $\Gamma_{if}$
between hyperfine states.
These parameters are deliberately
regarded separately from a
second set of parameters, 
\begin{align}
	\setTwo \defeq 
	\left\{ 
		\Omega_{\text{car},a,\uarr},
		\Omega_{\text{car},a,e},
		\Delta_{\text{car},a,\uarr},
		\Delta_{\text{blue},\darr,\uarr},
		a
	\right\}\,,
	\label{eq:setTwo}
\end{align}
consisting of the carrier Rabi
frequencies
and
detunings for both ground state
transitions
and a balance parameter $a$, which
shall become important during the optimization.
The carrier Rabi frequencies are directly
determined by the applied field strengths and
can be tuned over broad ranges.
The detunings $\Delta_{\text{car},a,\uarr}$
and $\Delta_{\text{blue},\darr,\uarr}$ should
be kept small to prevent off-resonant coupling
to additional motional modes.

\subsection{Optimization algorithm}

\begin{figure*}[tb]
	\includegraphics[scale=1]{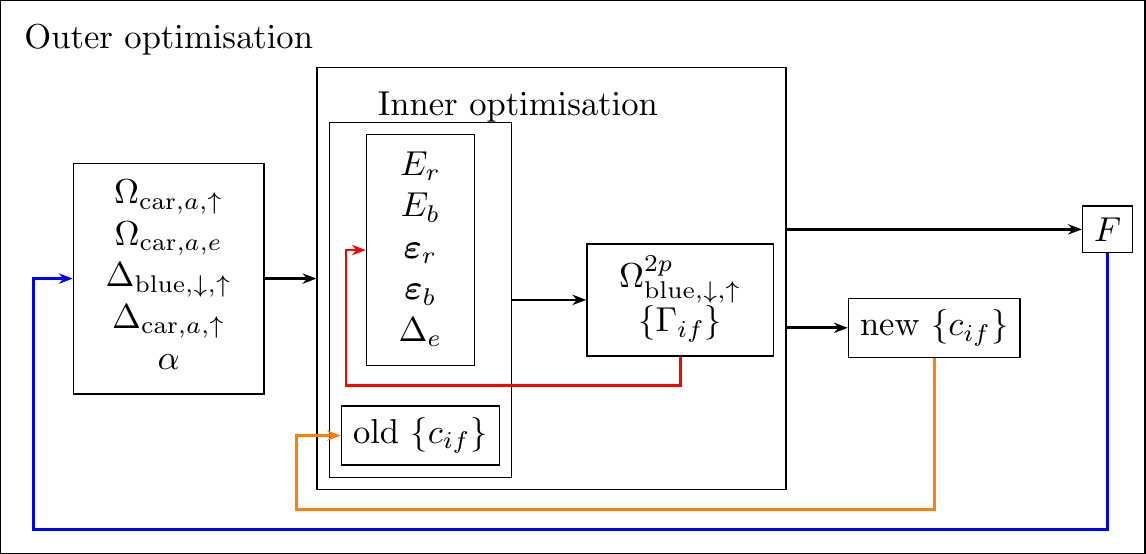}
	\caption{	
	\label{fig:algo}
	Schematic overview of the two-step parameter optimization algorithm
	for the two sets of parameters, $\setOne$ and $\setTwo$.
	The inner optimization (red loop) depends on $\setOne$
	and a set of weights $\left\{ c_{if} \right\}$.
	After the inner optimization, the old
	set of weights can be updated (orange loop)
	and the fidelity $F$ of the dynamics
	is optimized in an outer (blue) loop.	
	}
\end{figure*}
Our optimization algorithm, schematically depicted in \cref{fig:algo}, 
takes the approach of
optimizing the sets
introduced above in a two-step process.
Conceptually, the 
inner optimization over the first set
of parameters $\setOne$ incorporates the
dynamics indirectly and is encapsulated by
an outer optimization over the second
set of parameters $\setTwo$, maximizing the
actual fidelity $F$.
This strategy is motivated by the fact that  
determining
$\Omega_{\text{blue},\darr,\uarr}^{2p} = 
\Omega_{\text{blue},\darr,\uarr}^{2p}
(E_{r},E_{b},\bm{\varepsilon}_{r},\bm{\varepsilon}_{b},\Delta_{e})$
and $\left\{ \Gamma_{if} = \Gamma_{if}
(E_{r},E_{b},\bm{\varepsilon}_{r},\bm{\varepsilon}_{b},\Delta_{e})\right\}$
does not require explicit knowledge of the dynamics and is 
therefore computationally inexpensive.

The target functional of the inner step of the
optimization depends
on the field strengths $E_{r}$ and $E_{b}$,
polarizations $\bm{\varepsilon}_{r}$ and 
$\bm{\varepsilon}_{b}$ and excited state
detuning $\Delta_{e}$ and is defined as
\begin{align}
	J_{\text{inner}}[E_{r},E_{b},\bm{\varepsilon}_{r},\bm{\varepsilon}_{b},\Delta_{e}] = \sum_{if} c_{if}\Gamma_{if} - \alpha \Omega_{sid}^{2p}\eqnEndSpace.
	\label{eq:PolFunct}
\end{align}
Here, $\alpha$ is a balance parameter
which weights up the relative importance of maximizing
$\Omega_{\text{blue},\darr,\uarr}^{2p}$
versus minimising the sum $\sum_{if} c_{if} \Gamma_{if}$,
for a given set of weights $\left\{ c_{if} \right\}$.
If the set of weights $\left\{ c_{if} \right\}$ and $a$ are fixed,
the inner optimization can calculate 
$\Omega_{\text{blue},\darr,\uarr}^{2p}$
and $\left\{ \Gamma_{if} \right\}$ in terms
of $\setOne$, which are passed back to the
outer part of the optimization, once $J_{\text{inner}}$
is minimal.

\begin{figure}[tb]
	\includegraphics[scale=1]{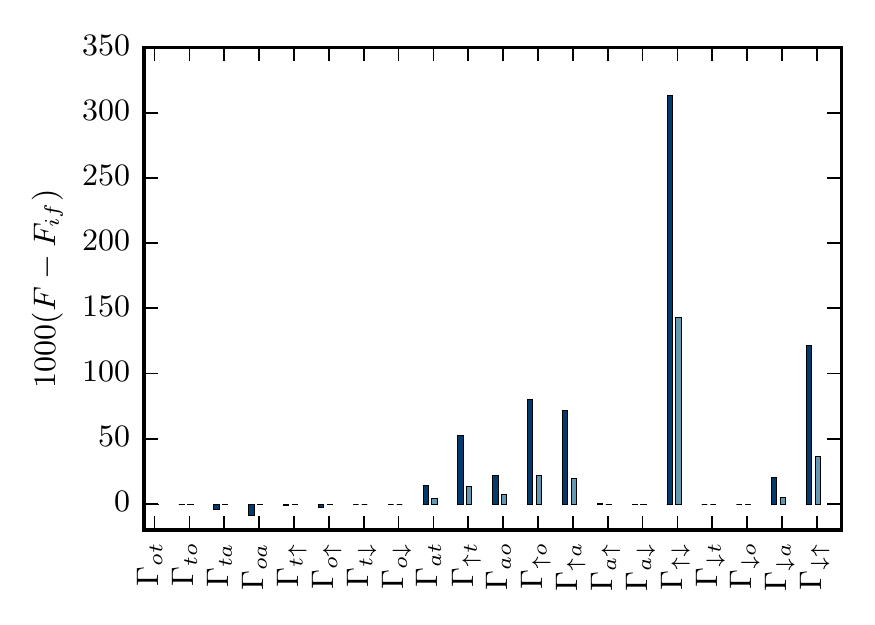}
	\caption{	
	\label{fig:weights} 
	Change in fidelity $F-F_{if}$ between unaltered
	dynamics and dynamics resulting from artificially 
	boosting a specific $\Gamma_{if}$. 
	The dark blue set
	of weights generated 
	in the first iteration is different from the pale
	blue set of weights generated after
	several updates.}
\end{figure}

The optimization of the set of parameters $\setOne$ 
requires a measurement of the effect
a change in each scattering rate $\Gamma_{if}$ has
on $F$, the overall fidelity of the dynamics.
This runs contrary to the usual practice of 
minimising the total scattering rate $\sum_{if} \Gamma_{if}$
between all pairs of ground state hyperfine levels.
Individually weighting each $\Gamma_{if}$ 
comes as a consequence of the 
observation that the leaking between each pair
of hyperfine ground states
affects the reached fidelity differently.
Most notably, transitions leading out
of the steady state $\Ket{S_{\darr\uarr}}$
and transitions leading out of the 
hyperfine subspace $\left\{ a,\darr,\uarr \right\}$
into the neighbouring states $\left\{ o,t \right\}$
have the largest negative effect on 
the fidelity.
Taking into account each individual leaking 
rate therefore offers the possibility
of strongly suppressing certain detrimental
$\Gamma_{if}$ by carefully tuning the polarization.
We encode the degree to which a certain $\Gamma_{if}$
affects the fidelity by running several simulations
where each individual rate
$\Gamma_{if}$ is artificially boosted 
by a factor of $10$, whilst keeping
all other rates fixed, resulting in a set of 
fidelities $\left\{F_{if}\right\}$. 
Observing the difference $F-F_{if}$,
between boosted and unaltered dynamics
leads to a set of weights, 
$\left\{ c_{if}\defeq 1000\left( F-F_{if} \right) \right\}$.

\Cref{fig:weights} shows two different
sets of weights at the beginning of the optimization and after a few updates. 
As discussed in \cite{lin2013dissipative}
the biggest scattering error is due to 
the qubit transition between $\Ket{\uarr}$
and $\Ket{\darr}$. 
$\Gamma_{\uarr\darr}$,
Along with transitions
leading out of the hyperfine subspace
into the neighbouring $\Ket{o}$ and $\Ket{t}$
levels, recieved the
largest weights $c_{if}$ for the
duration of the optimization.

A given set of weights can be used 
to optimize $\setOne$, leading to
the best possible
$\Omega_{\text{blue},\darr,\uarr}^{2p}$
and $\left\{ \Gamma_{if} \right\}$
with which to perform the dynamics.

The optimization of the second set 
of parameters, $\setTwo$, directly targets 
the fidelity $F$ of the dynamics, 
\begin{align}
	J_{\text{outer}}\left[ \Omega_{\text{car},a,\uarr}, \Omega_{\text{car},a,e},\Delta_{\text{blue},\darr,\uarr}, \Delta_{\text{car},a,\uarr},\alpha\right] \defeq 1 - F = \epsilon
	\label{eq:DynFunct}\,.
\end{align}
For each iteration of the outer optimization,
the inner optimization
over \cref{eq:PolFunct} leading to optimal
$\Omega_{\text{blue},\darr,\uarr}^{2p}$ and 
$\left\{ \Gamma_{if} \right\}$ is performed
using the set of weights $\left\{ c_{if} \right\}$
generated during the previous iteration 
(for the first iteration $c_{if}=1,~\forall i,f$).
After the inner optimization,
a new set of weights $\left\{ c_{if} \right\}$
is generated for the next iteration of
the outer optimization, as illustrated in \cref{fig:algo}.

This two-step optimization is 
easily generalised for arbitrary 
combinations of transitions, 
including the possibility for
multiple sideband transitions
between differentground state hyperfine
levels.
Optimization of multiple sideband
transitions follows the rule,
that the $j^{\text{th}}$
sideband transition
has its own set of polarizations
$\bm{\varepsilon}_{r}^{(j)}$ and
$\bm{\varepsilon}_{b}^{(j)}$, field
strengths $E_{r}^{(j)}$ and $E_{b}^{(j)}$,
balance parameter $\alpha^{(j)}$ and
excited state detuning $\Delta_{e}^{(j)}$
but
each contributes towards a set of
total scattering rates 
$\left\{ \Gamma_{if} = \sum_{j} \Gamma_{if}^{(j)} \right\}$.
Furthermore, all transitions except for
the repump transition have a detuning
$\Delta_{\text{type},i,f}$
and all carrier transitions have a
Rabi frequency $\Omega_{\text{car},i,f}$ to be optimized
directly, along
with the set of balance parameters
$\left\{ \alpha^{(j)} \right\}$, in the 
outer optimization.

\subsection{Result of optimization}
\label{ssect:origResult}
\begin{figure}[tb]
	\includegraphics[scale=1]{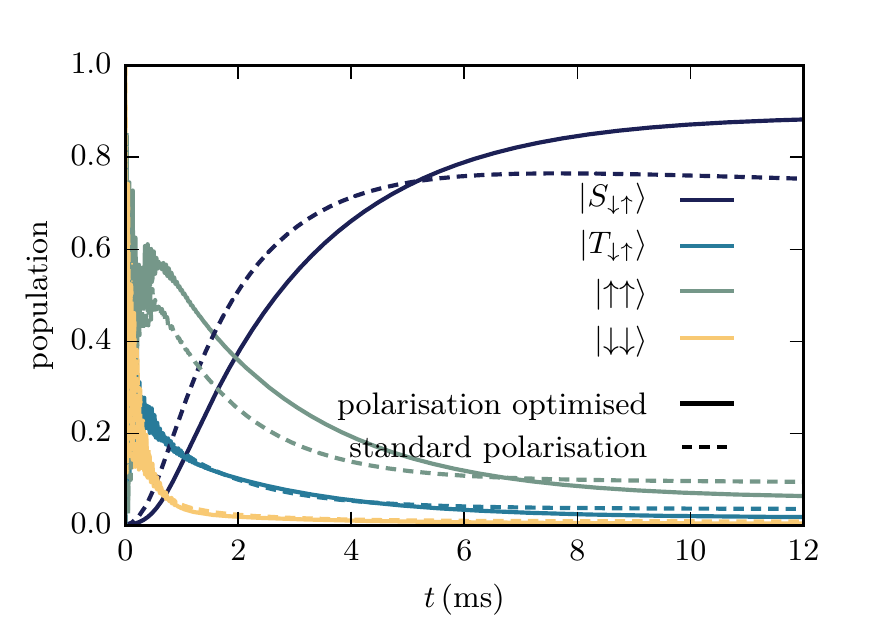}
	\caption{\label{fig:originalOptimisedDynamics} 
	Time-dependent population in the states
	$\Ket{\darr\darr}$, $\Ket{\uarr\uarr}$, 
	$\Ket{S_{\darr\uarr}}$ and $\Ket{T_{\darr\uarr}}$
	after summing over all motional levels.
	The graph compares the dynamics before (dashed lines)
	and after (solid lines) the combined optimization,
	leading to fidelities of $F=76\%$ and $88\%$ respectively.}
\end{figure}
\begin{table*}
	\centering
	\begin{tabular}{l | r| r| r| r| r| r| r| r| r| r}
		quantity & $E_{r}$ & $E_{b}$ & $\bm{\varepsilon}_{r}$ & 
			$\bm{\varepsilon}_{b}$ & 
			$\frac{\Omega_{\text{car},a,\uarr}}{2\pi}$ &
			$\frac{\Omega_{\text{blue},\darr,\uarr}^{2p}}{2\pi}$ &
			$\frac{\Omega_{\text{car},a,e}}{2\pi}$ &
			$\Delta_{\text{car},a,\uarr}$ & $\Delta_{\text{blue},\darr,\uarr}$ &
			$\Delta_{e}$ \\ \hline
      value & $\unit[7520]{\frac{V}{m}}$ & $\unit[7520]{\frac{V}{m}}$
      & $\left(0.162,0.987,0.000\right)$ & $\left(-0.870,-0.286,-0.403\right)$ 
      & $\unit[316]{Hz}$ & $\unit[7.65]{kHz}$ & $\unit[179]{kHz}$
      & $\unit[-46]{Hz}$ & $\unit[-44]{Hz}$ & $\unit[662]{GHz}$ \\
	\end{tabular}
	\caption{
		Optimized parameters when using the same transitions as in Ref.~\cite{lin2013dissipative},  
                leading to a fidelity of $F=88\%$, compared to $F=76\%$ in Ref.~\cite{lin2013dissipative}. 
		Both field strengths $E_{r}$ and $E_{b}$
		are limited to the maximum values of 
		$\unit[7520]{\frac{V}{m}}$.
		$\Omega_{\text{blue},\darr,\uarr}^{2p}$ is 
		determined by \cref{eq:SidebandRabifreq}, 
		$\Omega_{\text{car},a,e}$ leads to 
		$\gamma_{if}$ in 
		\cref{eq:GammaRepFirst,eq:GammaRepSecond}.
}
	\label{tab:originalOptimizedParams}
\end{table*}
All parameter optimizations have been performed
with the NLopt package~\cite{johnson2014nlopt}
using
the Subplex algorithm~\cite{rowan1990functional}.
While other optimization
methods could also be used 
in the outer and inner optimization loops, we have found these to converge well.
\Cref{fig:originalOptimisedDynamics} compares
the simulated dynamics of the system as described in
Ref.~\cite{lin2013dissipative} 
with the dynamics obtained after optimization.
The peak fidelity 
is increased from $F=76\%$ to
$F=88\%$. 
This is due to  a modified steady state, in which the populations
in $\Ket{T_{\darr\uarr}}$, $\Ket{\uarr\uarr}$
and $\Ket{\darr\darr}$ each are smaller 
than in the original scheme.
Furthermore, through optimization of
the $\Gamma_{if}$, a significant portion of the
population can be prevented from escaping
the ground state hyperfine subspace 
$\left\{ a,\darr,\uarr \right\}$, which
causes the prominent crest in the
$\Ket{S_{\darr\uarr}}$ population for the original scheme.
The optimized result was compared to
different realisations of randomly chosen
polarizations $\bm{\varepsilon}_{r}$ and
$\bm{\varepsilon}_{b}$, which leads to
dramatically varying peak fidelities
that can be as low as $F=10\%$ but are only
rarely in the vicinity of but never surpass
the peak fidelity reached by optimization.

The optimized values of the various parameters are 
reported in 
\cref{tab:originalOptimizedParams}.
After optimization, the two-photon 
sideband Rabi frequency $\Omega_{\text{blue},\darr,\uarr}^{2p}$
assumes a value of $2\pi \times \unit[7.65]{kHz}$,
which is very close to 
the rate $2\pi\times \unit[7.81]{kHz}$ reported in
Ref.~\cite{lin2013dissipative}.
The increase in fidelity can therefore
mainly be attributed to the
adjustments made to the polarization
$\bm{\varepsilon}_{r}$, $\bm{\varepsilon}_{b}$
and increase in excited state manifold detuning 
$\Delta_{e}$ from
$\unit[270]{GHz}$ to $\unit[662]{GHz}$, which
is feasible, see for example in Ref.~\cite{gaebler2016high}.
In other words, the outcome of the inner optimization is
a superior set of scattering rates
$\left\{ \Gamma_{if} \right\}$,
with the parameters of the outer optimization
adjusted to rebalance the system.
Compared to Ref.~\cite{lin2013dissipative}, in
which $\Omega_{\text{car},a,\uarr}=\unit[495]{Hz}$, 
the carrier Rabi frequency between $\Ket{a}$ and
$\Ket{\uarr}$ drops to $\unit[316]{Hz}$ after
optimization, thus further suppressing the
unwanted $\Ket{S_{\darr\uarr}}\leftrightarrow\Ket{S_{a\darr}}$
transition.
As the optimal fidelity
is approached, the detunings
$\Delta_{\text{car},a,\uarr}$ and
$\Delta_{\text{blue},\darr,\uarr}$
become negligibly small, indicating
that for this particular entanglement scheme, 
the shift out of resonance due to the
driven transitions is not much of a
factor.

Nevertheless, the achievable fidelity is inherently
limited in this entanglement scheme.
As demonstrated by 
\cref{eq:SidebandRabifreq,eq:Kramers,eq:aif}, 
even if the field strengths of
the lasers utilized for the stimulated Raman
sideband transition were unconstrained, 
a finite amount of leaking between hyperfine
states would remain present.
Limited field strengths of the sideband lasers
necessitate a trade-off between the error due
to leaking between hyperfine states and the
errors due to population trapping in $\Ket{\uarr\uarr}$
and the unfavourable transition between
$\Ket{S_{\darr\uarr}}$ and $\Ket{S_{a\darr}}$.

As such, the fidelity that can be reached with our optimized parameters 
falls short of the fidelity obtained by switching to the
stepwise scheme presented in Ref.~\cite{lin2013dissipative} which amounts to 
 $F=89.2\%$.
The stepwise scheme negates the error caused
by the unfavourable transition between 
$\Ket{S_{\darr\uarr}}$ and $\Ket{S_{a\uarr}}$
by temporally separating the 
ground state hyperfine transitions from the
application of the repumper and also the
sympathetic cooling.
This strategy ensures that population lost out
of $\Ket{S_{\darr\uarr}}$ into $\Ket{S_{a\uarr}}$
has nowhere to go and, if precisely timed,
is returned to $\Ket{S_{\darr\uarr}}$ after
a full Rabi cycle. 
Essentially, the stepwise scheme
lifts the requirement of
balancing the rates at which each
transition can be driven, thereby 
overcoming the limitations associated
with the time-continuous implementation.
In the following we will show that a continuously operated scheme can outperform both variants for entanglement generation of Ref.~\cite{lin2013dissipative} by exploiting a different combination of transitions. 

\section{Two-sideband scheme}
\label{sect:twosid} 
Alternatively to the original scheme presented in 
\cref{ssect:originalScheme}, steady-state entanglement
can be reached using other combinations of continuously
driven carrier and sideband
\beion-hyperfine transitions. 
We consider here a scheme that 
features two sideband transitions:
a blue sideband transition from $\Ket{\darr}$ to $\Ket{\uarr}$,
and a second, red sideband transition from $\Ket{\uarr}$ to $\Ket{a}$.
Note that we assume each sideband transition to be driven by its
own pair of stimulated Raman laser beams. It would also be possible to drive the two sideband transitions using only three beams. This simply requires
 proper choice of the correct
relative detunings.
In addition, and as in the original
scheme, a repump transition between
$\Ket{a}$ and $\Ket{e}$ is driven.
In order for all states in the hyperfine subspace to
be connected to the target state $\Ket{S_{\darr\uarr}}$,
a carrier transition between 
$\Ket{\darr}$ and $\Ket{\uarr}$ is included as well.
This choice is similar to the combination of transitions utilized 
for the entanglement of two $^{40}\text{Ca}^{+}$ 
ions in Ref.~\cite{bentley2014detection}. It  offers 
numerous advantages over the original scheme as detailed below.

\subsection{Entanglement mechanism and optimization parameters}
\label{ss:twosidEntMech}
\begin{figure*}[tb]
	\includegraphics[scale=0.8]{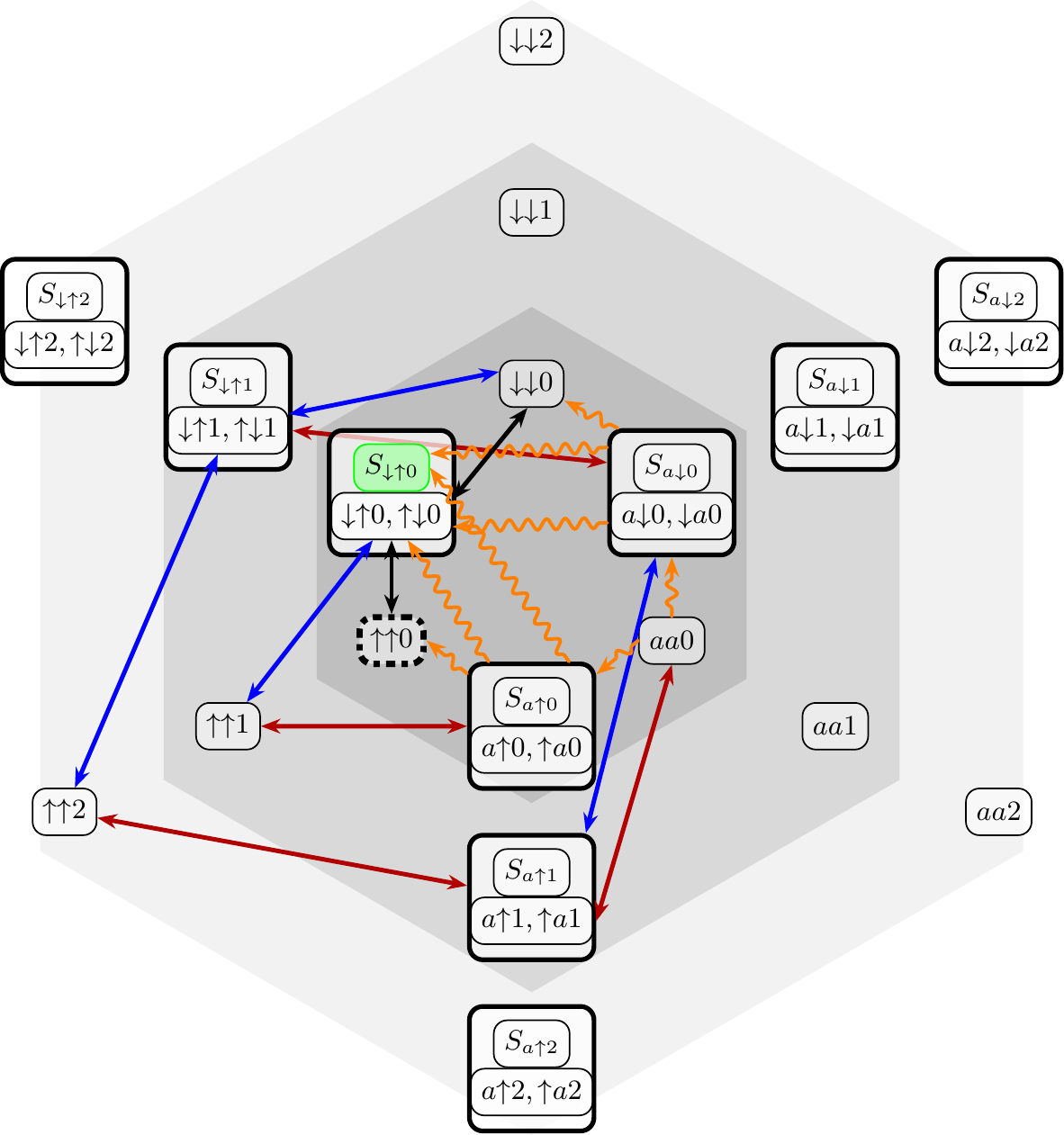}
	\caption{\label{fig:twoSBMechanism} 
	Graphical overview of transitions 
	for the two-sideband transition scheme.
	As in the original scheme, only the most
	critical transitions for entanglement
	are shown and only states of the 
	hyperfine subspace $\left\{ a,\darr,\uarr \right\}$
	and the \modeOne\,motional mode are displayed.
	Carrier transitions between $\Ket{\darr}$ and
	$\Ket{\uarr}$ at rate $\Omega_{\text{car},\darr,\uarr}$
	appear as black double-headed
	arrows.
	The blue $\Ket{\darr} \rightarrow \Ket{\uarr}$
	sideband transitions act at a rate of
	$\Omega_{\text{blue},\darr,\uarr}^{2p}$
	and appear as blue double headed arrows,
	whilst the red $\Ket{\uarr}\rightarrow \Ket{a}$
	sideband transitions acts at a rate
	$\Omega_{\text{red},\uarr,a}^{2p}$ and appear as
	red double headed arrows.
	Effective decay out of $\Ket{a}$ into the
	hyperfine subspace appears as orange snaking
	lines.
	Sympathetic cooling is no longer incorporated
	into the mechanism and heating of the motional 
	mode and leaking between hyperfine states
	are not shown.} 
\end{figure*}
\Cref{fig:twoSBMechanism} illustrates the entanglement
mechanism for this new combination of
transitions.
Crucially, the unfavourable transition
between $\Ket{S_{\darr\uarr}}\otimes\Ket{0}$ and 
$\Ket{S_{a \uarr}}\otimes\Ket{0}$ due to the carrier
connecting $\Ket{a}$ and $\Ket{\uarr}$
in the original scheme has been eliminated.
Instead, the red sideband transition from $\Ket{\uarr}$
to $\Ket{a}$ leads from
$\Ket{S_{\darr\uarr}}\otimes\Ket{n_{\modeOne}}$
to $\Ket{S_{a\darr}}\otimes\Ket{n_{\modeOne}-1}$
only when $n_{\modeOne} > 0$.
Consequently, for this combination of transitions,
in the absence of leakage between
hyperfine states and heating, 
$\Ket{S_{\darr\uarr}}\otimes\Ket{n_{\modeOne}=0}$ alone
is the steady state of the dynamics.
In the presence of heating, population in 
$\Ket{S_{\darr\uarr}}\otimes\Ket{0}$ can
only escape due to an excitation of the
utilized vibrational mode $\modeOne$
followed by a sideband transition from
$\Ket{S_{\darr\uarr}}\otimes\Ket{n_{\modeOne}}$
to $\Ket{S_{a\darr}}\otimes\Ket{n_{\modeOne}-1}$.
Population in $\Ket{S_{a\darr}}\otimes\Ket{n_{\modeOne}-1}$
can take multiple branching paths,
all of which
eventually lead back to 
$\Ket{S_{\darr\uarr}}$.
As such, in contrast to the 
original scheme, which relies on
sympathetic cooling, this particular combination of
transitions inherently cools the utilized
mode \modeOne$\,$  of the system
during entanglement generation.

Without the need for sympathetic cooling, the 
\mgion ions can be removed. 
This leads not only to
a simplification of the experiment but also reduces the
number of motional modes of the ionic
crystal. It thus effectively eliminates
the error due to off-resonant coupling
to $\modeTwo$, given by
\cref{eq:SBORtrans} in the original scheme. 
As described in \cref{sect:Optim},
in the original scheme
the carrier Rabi frequencies 
$\Omega_{\text{car},a,\uarr}$ and 
$\Omega_{\text{car},a,e}$,
which determine the rate of
effective decay out of $\Ket{a}$,
are limited by the maximum attainable
$\Omega_{\text{blue},\darr,\uarr}^{2p}$.
In contrast, in the current scheme
the carrier Rabi frequencies 
$\Omega_{\text{car},\darr,\uarr}$ and
$\Omega_{\text{car},a,e}$ can be  
increased significantly, without
causing losses out of the target state
and population trapping in 
$\Ket{\uarr\uarr}$.
By driving an additional sideband
transition, the graph of states
in \cref{fig:twoSBMechanism}
is more connected, permitting
population to reach $\Ket{S_{\darr\uarr}}$
by additional paths.
Comparing the graphs shown in 
\cref{fig:origMechanism,fig:twoSBMechanism},
the combined effect of additional paths
into the target state and the increase
in $\Omega_{\text{car},a,e}$ which results in larger 
effective decay rates 
$\left\{\gamma_{a,f} \propto \Omega_{\text{car},a,e}^{2} \right\}$
should lead to much faster entanglement preparation.

Optimization of the field strengths
and polarizations for the two-sideband
scheme has been carried out according to the same
principle as described in
\cref{sect:Optim}, with the slight
complication of having to address
additional degrees of freedom.
In the specific case of the two-sideband
combination, the corresponding form
of the target functional for the polarization optimization,
\cref{eq:PolFunct}, becomes
\begin{align}
	J_{\text{inner}}[&E_{b}^{(1)},E_{r}^{(1)},\varepsilon_{b}^{(1)},
	\varepsilon_{r}^{(1)},\Delta_{e}^{(1)}, \nonumber \\
	& E_{b}^{(2)},E_{r}^{(2)},\varepsilon_{b}^{(2)},
	\varepsilon_{r}^{(2)},\Delta_{e}^{(2)}] \nonumber\\
	= &\sum_{if} c_{if} \Gamma_{if} - \alpha^{(1)} \Omega_{\text{blue},\darr,\uarr}^{2p~(1)} - \alpha^{(2)} \Omega_{\text{red},\uarr,a}^{2p~(2)}\eqnEndSpace.&
	\label{eq:InnerFunctTwoSid}  
\end{align}
As in the original scheme
and described in detail in \cref{sect:Optim}, 
optimization of the
polarization can be accomplished
without having to simulate the
dynamics in each iteration.
A single inner optimization step determines both
$\Omega_{\text{blue},\darr,\uarr}^{2p~(1)}
=\Omega_{\text{blue},\darr,\uarr}^{2p~(1)}
\left(E_{r}^{(1)},E_{b}^{(1)},\bm{\varepsilon}_{r}^{(1)},
\bm{\varepsilon}_{b}^{(1)},\Delta_{e}^{(1)}\right)
$
and
$\Omega_{\text{red},\uarr,a}^{2p~(2)}
=\Omega_{\text{red},\uarr,a}^{2p~(2)}
\left(E_{r}^{(2)},E_{b}^{(2)},\bm{\varepsilon}_{r}^{(2)},
\bm{\varepsilon}_{b}^{(2)},\Delta_{e}^{(2)}\right)
$,
in addition to 
$\big\{ \Gamma_{if} = \Gamma_{if}^{(1)}
\left(E_{r}^{(1)},E_{b}^{(1)},\bm{\varepsilon}_{r}^{(1)},
\bm{\varepsilon}_{b}^{(1)},\Delta_{e}^{(1)}\right) 
+ \Gamma_{if}^{(2)} 
\left(E_{r}^{(2)},E_{b}^{(2)},\bm{\varepsilon}_{r}^{(2)},
\bm{\varepsilon}_{b}^{(2)},\Delta_{e}^{(2)}\right)
\big\}$,
the set of scattering rates due to each
sideband transition.
As in \ref{sect:Optim}, the outer optimization
is performed directly on the fidelity $F$ of the dynamics, 
\begin{align}
	J_{\text{outer}}[&\Omega_{\text{car},\darr,\uarr},\Omega_{\text{car},a,e},\Delta_{\text{car},\darr,\uarr}, 
		\nonumber \\
		&\Delta_{\text{blue},\darr,\uarr},
	\Delta_{\text{red},\uarr,a}, \alpha^{(1)}, \alpha^{(2)}] \defeq 1 - F = \epsilon\,.
	\label{eq:OuterFunctTwoSid}
\end{align}
The set 
$\setTwo$ now consists of
the carrier Rabi frequencies 
$\Omega_{\text{car},\darr,\uarr}$,
$\Omega_{\text{car},a,e}$, 
the detunings
$\Delta_{\text{car},\darr,\uarr}$,
$\Delta_{\text{blue},\darr,\uarr}^{(1)}$
and $\Delta_{\text{red},\uarr,a}^{(2)}$,
of the microwave carrier, first and second sideband
transitions, and 
the weights
$\alpha^{(1)}$ and 
$\alpha^{(2)}$. 
Since the scheme now involves a second sideband combination,
an additional weight is required to balance
the maximization of its two-photon sideband 
Rabi frequency against the sideband photon
scattering rates in \cref{eq:InnerFunctTwoSid}.
In order to make sure that
both Rabi frequencies are
maximized without one
dominating the other, however,
the left hand side of 
\cref{eq:InnerFunctTwoSid} can be
modified slightly, such that
\begin{align} 
	\tilde{J}_{\text{inner}}[&E_{b}^{(1)},E_{r}^{(1)},\varepsilon_{b}^{(1)},
		\varepsilon_{r}^{(1)},\Delta_{e}^{(1)}, \nonumber \\
	& E_{b}^{(2)},E_{r}^{(2)},\varepsilon_{b}^{(2)},
	\varepsilon_{r}^{(2)},\Delta_{e}^{(2)}]  \nonumber\\
 	& = \sum_{if} c_{if}\Gamma_{if} - \alpha 
		\left( \Omega_{\text{blue},\darr,\uarr}^{2p~(1)} 
		+\Omega_{\text{red},\uarr,a}^{2p~(2)} \right) \nonumber\\ 
		&~~ + \beta \left| \Omega_{\text{blue},\darr,\uarr}^{2p~(1)} 
		-\Omega_{\text{red},\uarr,a}^{2p~(2)} \right| \eqnEndSpace,
	\label{eq:ModifiedInnerFunctTwoSid}
\end{align}
where $\alpha$ now balances the
maximization of the sum of
two-photon Rabi frequencies
against the $\Gamma_{if}$, whilst
$\beta$ 
is a parameter controlling how
strictly the two-photon sideband
Rabi frequencies should be matched.
For simplicity it is assumed that
$E_{r}^{(1)} = E_{r}^{(2)}$ and
$E_{b}^{(1)} = E_{b}^{(2)}$ and
that each field strength is
limited to the maximum value 
allowed during the optimization
of the original scheme.

\begin{figure}[tb]
	\includegraphics[scale=1]{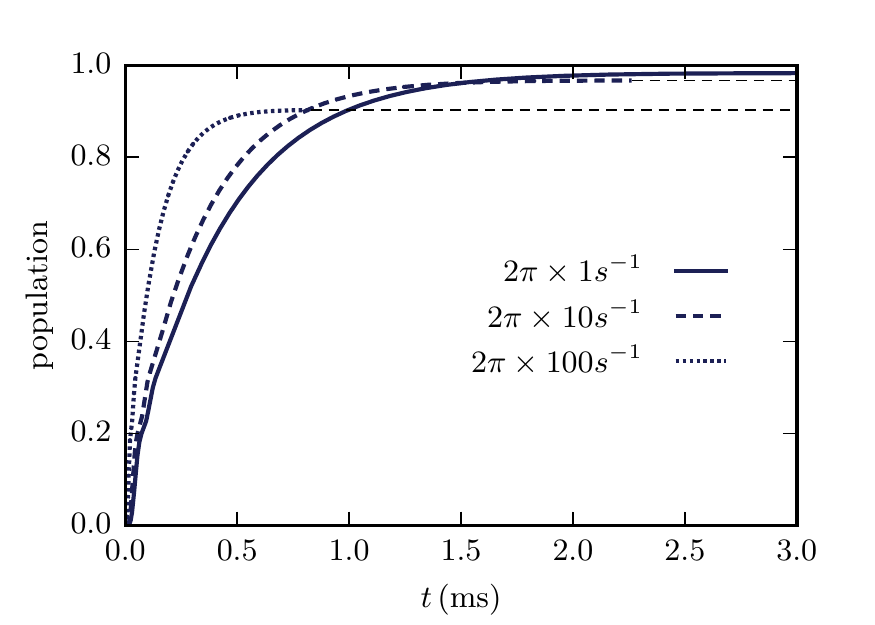}
	\caption{\label{fig:twoSBDynamics}
	Population of the target state 
	$\Ket{S_{\darr\uarr}}$ over time obtained with 
        by optimization of the peak fidelity at final times $T$, for
	the two-sideband scheme with different assumed heating rates.
	The plot shows curves for the heating rates $\kappa_{\text{h}}$
	of $2\pi\times\unit[1]{\rate}$ (solid line) leading to  a fidelity of 
	$F=98.3\%$,
	$2\pi\times\unit[10]{\rate}$ (dashed line) leading to $F=96.7\%$ and
	$2\pi\times\unit[100]{\rate}$ (dotted line) leading to $F=90.3\%$.
}
\end{figure}
\begin{table*}[tb]
	\centering
	\begin{tabular}{l |l r r r }
		parameter & $\kappa_{h}=\quad $ & $2\pi\times\unit[1]{\rate}$ & $2\pi\times\unit[10]{\rate}$ & $2\pi\times\unit[100]{\rate}$ \\ \hline
		$E_{r}$ && 
			$\unit[7520]{\frac{V}{m}}$ & 
			$\unit[7520]{\frac{V}{m}}$ & 
			$\unit[7520]{\frac{V}{m}}$  \\ 
		$E_{b}$ && 
			$\unit[7520]{\frac{V}{m}}$ & 
			$\unit[7520]{\frac{V}{m}}$ & 
			$\unit[7520]{\frac{V}{m}}$  \\
		$\bm{\varepsilon}_{r}^{(1)}$ && 
			$\left( -0.752,-0.220,-0.621 \right)$ & 
			$\left( -0.620, -0.500, -0.605 \right)$& 
			$\left( -0.741, -0.338, -0.581 \right)$ \\
		$\bm{\varepsilon}_{b}^{(1)}$ &&
			$\left( 0.440, 0.759, 0.480 \right)$ & 
			$\left(0.536, 0.644, 0.545 \right)$ & 
			$\left(0.408,0.802, 0.435 \right)$ \\
		$\bm{\varepsilon}_{r}^{(2)}$ &&
			$\left( -0.413,-0.204,-0.888 \right)$ & 
			$\left( -0.453, -0.854, -0.257 \right)$& 
			$\left( -0.479, -0.824, -0.303 \right)$ \\
		$\bm{\varepsilon}_{b}^{(2)}$ && 
			$\left( -0.415, -0.883, -0.218 \right)$ & 
			$\left( -0.451, -0.250, -0.857\right)$& 
			$\left(  0.493,  0.261, 0.830 \right)$ \\
		$\Omega_{\text{car},\darr,\uarr}$ && 
			$2\pi\times\unit[2.24]{kHz}$ & 
			$2\pi\times\unit[2.91]{kHz}$ &
			$2\pi\times\unit[6.67]{kHz}$\\
		$\Omega_{\text{blue},\darr,\uarr}^{2p~(1)}$ && 
			$2\pi\times\unit[4.96]{kHz}$& 
			$2\pi\times\unit[6.47]{kHz}$& 
			$2\pi\times\unit[14.92]{kHz}$ \\ 
		$\Omega_{\text{red},\uarr,a}^{2p~(2)}$ && 
			$2\pi\times\unit[4.96]{kHz}$ & 
			$2\pi\times\unit[6.47]{kHz}$& 
			$2\pi\times\unit[14.92]{kHz}$ \\ 
		$\Omega_{\text{car},a,e}$ && 
			$2\pi\times\unit[691]{kHz}$ & 
			$2\pi\times\unit[802]{kHz}$ & 
			$2\pi\times \unit[1233]{kHz}$ \\ 
		$\Delta_{e}^{(1)}$ && 
			$\unit[624]{GHz}$ & $\unit[245]{GHz}$ & $\unit[318]{GHz}$ \\
		$\Delta_{e}^{(2)}$ && 
			$\unit[464]{GHz}$ & $\unit[372]{GHz}$ & $\unit[206]{GHz}$ \\
	\end{tabular}
	\caption{
		Optimized parameters for the two-sideband scheme.
		As for the original scheme, each field strength
		is limited to a maximum value of 
		$\unit[7520]{V/m}$.
		$\Omega_{\text{blue},\darr,\uarr}^{2p~(1)}$ and
		$\Omega_{\text{red},\uarr,a}^{2p~(2)}$ are both
		determined by \cref{eq:SidebandRabifreq}
		with individual polarizations
		$\bm{\varepsilon}_{r}^{(1)}$,
		$\bm{\varepsilon}_{b}^{(1)}$,
		$\bm{\varepsilon}_{r}^{(2)}$ and
		$\bm{\varepsilon}_{b}^{(2)}$.
		Optimization of the two-sideband scheme
		leads to fidelities of $F=98.3\%, 96.7\%$ 
		and $90.3\%$ for heating rates of 
		$\kappa_{\text{h}} = 2\pi\times\unit[1]{\rate},
		2\pi\times\unit[10]{\rate}$ 
		and $2\pi\times\unit[100]{\rate}$, respectively.
		The sideband detunings $\Delta_{e}^{(1)}$ and
		$\Delta_{e}^{(2)}$ are defined as in \cref{ssect:Ham}
		with the same assumed fine structure splitting 
		$f_{P}=\unit[197.2]{GHz}$}.
	\label{tab:twoSBOptimizedParams}
\end{table*}
In the absence of sympathetic cooling,
the primary source of heating,
caused by spontaneous emission 
during the stimulated Raman sideband
transition driven on the \mgion ions,
is eliminated.
The remaining sources of heating are
photon recoil from the spontaneous
emission out of $\Ket{e}$ after repumping
and electric field noise associated 
with the ion trap \cite{lin2013dissipative}.
Since the
heating rate
influences the system dynamics
and therefore the obtained fidelity,
the result of the optimization depends
on the specific heating rate assumed,
which can vary, depending on the
motional mode utilized for the sideband
transition.

\subsection{Influence of trap heating rates}

\Cref{fig:twoSBDynamics} compares the
reached peak fidelity for different
values of the
heating rate
$\kappa_{\text{h}}$ 
of 
vibrational mode \modeOne.
For an assumed heating rate of
$\kappa_{\text{h}} = 2\pi \times \unit[1]{\rate}$,
optimization leads to a peak fidelity of $F=98.3\%$, whilst
$\kappa_{\text{h}} = 2\pi \times \unit[10]{\rate}$
is a more realistic heating rate for modern traps
leading to a peak fidelity of $F=96.7\%$.
Finally, when $\kappa_{\text{h}} = 2\pi \times \unit[100]{\rate}$,
the peak fidelity is reduced to $F=90.3\%$. 
For each considered heating rate,
the parameters leading to optimal entanglement
are listed in \Cref{tab:twoSBOptimizedParams}.
As the heating rate $\kappa_{\text{h}}$
is increased, recovering population
lost from the target
state $\Ket{S_{\darr\uarr}}$
requires an increase in
the Rabi frequencies of all driven
transitions.
For all reported heating rates,
however, the ratios 
$\Omega_{\text{car},\darr,\uarr} \propto 
\Omega_{\text{blue},\darr,\uarr}^{2p~(1)} \approx 
\Omega_{\text{red},\uarr,a}^{2p~(2)} \propto 
\Omega_{\text{car},a,e}^{2}$
remain approximately constant.
Here, the repumper Rabi frequency $\Omega_{\text{car},a,e}$
enters squared, since the effective decay rates
in \cref{eq:EffectiveRates}
are proportional to $\Omega_{\text{car},a,e}^{2}$.
This observation can be understood, 
since the target state should be reachable as
directly as possible from any given state.
Scaling all transition rates equally
is necessary in order to prevent the flow of
population from being bottlenecked throughout the 
entanglement generation.
The optimized peak fidelities are again
significantly higher than the average fidelity
of $F\approx 0.4$ (or $F\approx 0.5$ with the 
fixed scaling of Rabi frequencies mentioned above) 
obtained from simulating the dynamics with
random polarizations 
$\bm{\varepsilon}_{r}^{(1)}, \bm{\varepsilon}_{b}^{(1)},
\bm{\varepsilon}_{r}^{(2)}$ and $\bm{\varepsilon}_{b}^{(2)}$
of the sideband laser beams and assuming 
$\kappa_{\text{h}}=2\pi\times\unit[1]{\rate}$.
For the two-sideband scheme it is much more difficult
to randomly select a near-optimal polarization, due 
to the increased number of degrees of freedom, which
also causes the peak fidelity to strongly vary depending
on the polarization.
Furthermore, an optimization to minimize the time taken 
to reach a target state population
of $F=85\%$ was performed for the heating
rates $\kappa_\text{h}\in 
\left\{2\pi\times \unit[1]{\rate},2\pi\times\unit[10]{\rate},2\pi\times\unit[100]{\rate} \right\}$,
leading to a preparation time of
$t\approx \unit[0.3]{ms}$ for all 
assumed heating rates.

For each increase in the heating rate, 
the optimization results 
in a different set of polarizations 
$\bm{\varepsilon}_{r}^{(1)}$, 
$\bm{\varepsilon}_{b}^{(1)}$, 
$\bm{\varepsilon}_{r}^{(2)}$ and 
$\bm{\varepsilon}_{b}^{(2)}$.
As the heating rate is increased, 
the minimization of leakage rates 
$\left\{\Gamma_{if}\right\}$ becomes less important.
A given $\Gamma_{if}$ is determined
by the polarization of each stimulated Raman laser beam
and scales with the
squared magnitude of the field strength $|E|^{2}$
whilst scaling inversely with the squared detuning
from the excited state $\Delta_{e}$ 
(\cref{eq:Kramers,eq:aif}) of the considered laser beam.
Instead, the maximization of the
two-photon stimulated Raman
sideband transition rates 
$\Omega_{\text{blue},\darr,\uarr}^{2p~(1)}$ and
$\Omega_{\text{red},\uarr,a}^{2p~(2)}$ (\cref{eq:SidebandRabifreq})
is prioritised.
The two-photon stimulated Raman sideband 
transition Rabi frequencies depend on
the polarizations $\bm{\varepsilon}_{r}$ and 
$\bm{\varepsilon}_{b}$ of both laser beams, 
the product of field strengths $E_{r}E_{b}$
and the detuning of both stimulated Raman laser beams
from the excited state $\Delta_{e}$.
Larger sideband two-photon Rabi frequencies 
ensure that population can flow back into 
$\Ket{S_{\darr\uarr}}\otimes\Ket{0}$
much faster than the heating can allow it to escape.

Increasing all of the transition rates
has the side effect of speeding up the entanglement
but limits the attainable fidelity, with an increased
error due to population leaking outside of the 
hyperfine subspace $\left\{ a,\darr,\uarr \right\}$.
The behaviour of the $\Delta_{e}^{(1)}$
and $\Delta_{e}^{(2)}$ is 
non-monotonic and appears to be strongly dependent on the
particular polarization profile.
As in the original scheme, 
each of the $\Delta_{\text{car},\darr,\uarr}$,
$\Delta_{\text{blue},\darr,\uarr}$ and
$\Delta_{\text{red},\uarr,a}$ becomes
smaller as the optimal fidelity is reached.

The error due to heating can only be
reduced by increasing the flow
of population into the target state
$\Ket{S_{\darr\uarr}}\otimes\Ket{0}$,
since there is no straightforward
way to compensate for heating.
This comes at the cost of 
increasing 
$\sum_{if} c_{if} \Gamma_{if}$ and 
thus the error due to leakage between the
hyperfine states, as explained above.
Assuming optimal polarization and
balancing of the driven rates,
the only way to reduce
one error without compounding
the other error
is by increasing the maximum field strengths $E_{r}^{(1)}$,
$E_{b}^{(1)}$,$E_{r}^{(2)}$ and $E_{b}^{(2)}$. This explains why the
field strengths take their maximal allowed value in
\Cref{tab:twoSBOptimizedParams}. 

\subsection{Comparison to the original scheme}

The two-sideband scheme  represents a promising alternative
to the original scheme even with optimized parameters, as
discussed in \cref{sect:Optim}.
In terms of fidelity, the two-sideband 
scheme outperforms the original one, 
regardless of the assumed heating
rate $\kappa_{\text{h}}$.
Even in the worst case considered, with 
$\kappa_{\text{h}} = 2\pi\times\unit[100]{\rate}$,
the resulting error is under $10\%$ after optimization.
In comparison, the previously best
fidelity, reached by the stepwise scheme 
in Ref.~\cite{lin2013dissipative}, corresponds to an error of about 
$11\%$.
The corresponding errors for the original scheme in \cref{sect:Optim}
are slightly larger for the polarization optimized case and
two and a half times as large for the non-optimized case.
In terms of speed, the two-sideband scheme
outperforms the original scheme.
Given traps with sufficiently small heating rates,
entangling speed can be sacrificed in order to maximize fidelity.
The lowest regarded heating rate
$\kappa_{\text{h}}=2\pi\times\unit[1]{\rate}$,
can be optimized over $\unit[3]{ms}$, attaining a
fidelity of $F=98.3\%$, or optimized over $\unit[6]{ms}$,
in order to increase the fidelity to $F=98.7\%$.
In contrast, the original scheme peaks after
approximately $\unit[6]{ms}$ but at the much lower fidelity $F=76\%$.

To summarize, when considering the experimental modifications 
necessary to go from the protocol 
in Ref.~\cite{lin2013dissipative} to
the two-sideband scheme, the overall complexity
is reduced.
Instead of a four ion setup consisting of two \beion and
two \mgion ions, with their respective sympathetic cooling
laser beams, now only the two \beion
to be entangled need to be trapped without sympathetic cooling laser beams.
Given sufficient power, the four laser 
beams required for both of the stimulated
Raman sideband transitions can all be derived from the
same $\unit[313]{nm}$ laser and frequency shifted using
acousto-optic modulators.
The only further complication is the ability to 
independently manipulate the polarization
of each individual stimulated Raman sideband
transition laser beam.
One may wonder of course how sensitive the Bell state fidelity is with respect to small deviations from the optimized polarizations. We have found 
fluctuations in the polarization components of up to $5\%$ to only have a neglible effect on the entanglement error, whilst fluctuations above $10\%$ will noticeably reduce the fidelity.

\subsection{Fundamental performance bound}

Given the superior performance of the two-sideband scheme compared to
the original protocol~\cite{lin2013dissipative}, one may wonder
whether there are ultimate limits to the fidelity of a Bell state
realized in this way. 
There are two main sources of error that limit the fidelities in this
dissipative state preparation scheme---anomalous heating and spontaneous
emission. As discussed above, the obtainable fidelity is determined by a
trade-off between utilizing fast enough sideband transitions in order
to beat trap heating, and minimization of the spontaneous
emission rates associated with the sideband transitions.
While anomalous heating can in principle be made
arbitrarily small by improving the ion trap, undesired spontaneous
emission is an inherent and unavoidable loss mechanism accompanying
the desired spontaneous emission at the core of the dissipative state
preparation. In order to explore the 
fundamental performance bound posed by spontaneous emission, we assume
a realistic trap with $\kappa_{h}=2\pi\times\unit[10]{\rate}$, a close to perfect 
trap, with $\kappa_{h}=2\pi\times\unit[1]{\rate}$
or no heating at all ($\kappa_{h}=\unit[0]{\rate}$), and investigate how much
laser power is needed to achieve a certain fidelity, or error. 

In the absence of all heating, the optimization will favor slow
sideband transitions that are detuned far below the
$P_{\nicefrac{1}{2}}$ and $P_{\nicefrac{3}{2}}$ levels with laser
beams polarized such that there is minimal spontaneous 
emission. Identifying the conditions under which it is possible to reach
Bell state fidelities of $F=99.9\%$ or even $F=99.99\%$
allows us to benchmark the performance of the current dissipative scheme.
For comparison, Ref.~\cite{ozeri2007errors} 
examines the dependence 
of fidelity on laser power for gate-based 
entanglement creation for various ion species.
Of all observed ion species, 
the gate error of \beion entanglement 
was lowest for a given power $P$, related to
the laser field strengh $E$ by
\begin{align}
	P = \frac{\pi}{4} E^{2}w_{0}^{2}c\epsilon_{0}\,\,,
	\label{eq:power}
\end{align}
where $w_{0}$ is the laser beam waist, 
$c$ the speed of light and  $\epsilon_{0}$ the vacuum
permittivity~\cite{ozeri2007errors}. We assume here an (idealized)
beam waist of $w_{0}=\unit[20]{\mu m}$, to directly compare
to Ref.~\cite{ozeri2007errors}.

During optimization, the highest regarded threshold, $F=99.99\%$
was reached after $\unit[0.33]{ms}$  using field strengths of 
$E_{r/b}\approx \unit[752]{kV/m}$ per beam and detunings
up to $\unit[25]{THz}$.
For the sake of comparison with Ref.~\cite{ozeri2007errors}, and
for the case of negligible heating, the timescale in the master equation 
	\eqref{eq:QME} can be changed, $t \rightarrow \tau= \frac{t}{\chi}$.
In order to match the same entangling speed and duration of $\unit[10]{\mu s}$ as reported in Ref.~\cite{ozeri2007errors}, we require $\unit[4.4]{MV/m}$ per beam,
corresponding to a total power of $4\times\unit[16]{W}$ at the same 
detuning.
For this very fast entanglement, the negative effects of heating are limited,
leading to errors of $\epsilon = 6.5\times 10^{-5}$, 
$\epsilon=1.0\times 10 ^{-4}$ and
$\epsilon=4.47\times 10^{-4}$ for heating rates of $\kappa_h \in \left\{2\pi\times\unit[1]{\rate},
2\pi\times\unit[10]{\rate}, 2\pi\times\unit[100]{\rate}\right\}$, respectively.

If we fix the available field strength to the value 
$E_{r/b}\approx\unit[200]{kV}$, corresponding to a total power 
of $4\times \unit[36]{mW}$,
as reported in Ref.~\cite{ozeri2007errors}, the target fidelity is reached after  $\unit[4.6]{ms}$.
Again, despite the much lower field strengths, the detuning remains unchanged.
It should be noted here, that an entangling duration of  $\unit[4.6]{ms}$ is 
still faster than that of the original entanglement scheme~\cite{lin2013dissipative}.
At this lower extreme in field strengths, the effects of 
heating are more noticeable, since heating is allowed to act
for almost $500$ times longer relative to the $\unit[10]{\mu s}$ case.

\begin{figure}
	\includegraphics{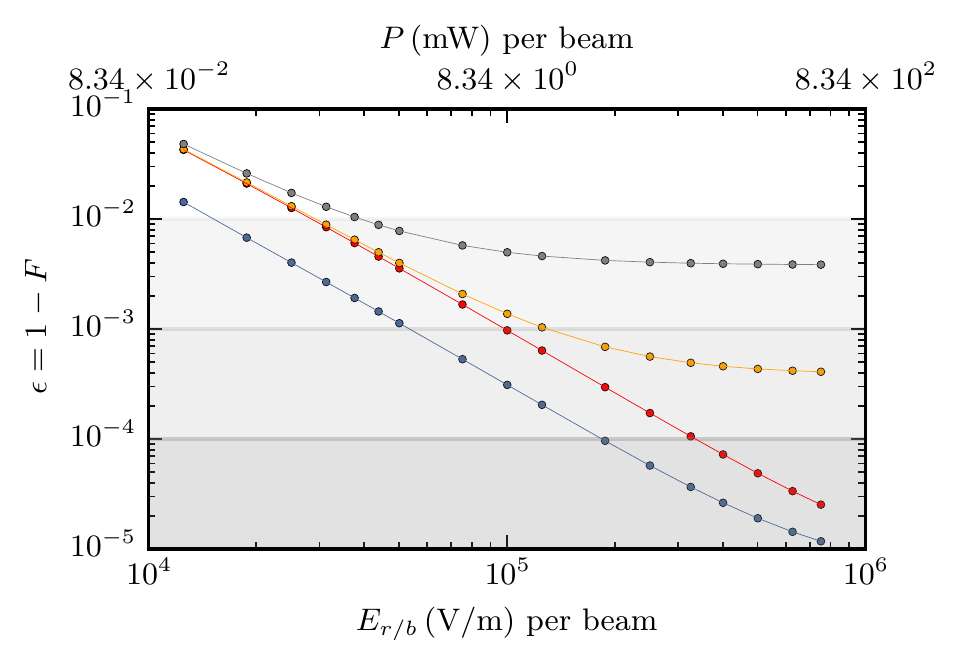}
	\caption{Bell state error  $\epsilon=1-F$
          as a function of the sideband laser beam strengths $E_{r/b}$
          allowed during optimization.
          The red (orange and grey) points correspond to an ion trap with
          $\kappa_{h}=0$ 
          ($\kappa_{h}=2\pi\times\unit[1]{\rate}$ and
          $\kappa_{h}=2\pi\times\unit[10]{\rate}$, respectively).
	  The blue points correspond to a three-beam configuration 
		  and $\kappa_{h}=0$.
          For all points, the detuning and carrier transition field strengths
          are chosen such that the fidelity peaks after an entangling duration
          of $\unit[1]{ms}$.
          For zero heating, attaining a maximal error of 
          $\epsilon < 0.001$ requires field strengths of $\unit[100]{kV/m}$, 
          or a combined power of $4\times \unit[8.3]{mW}$ going into a 
          20$\,\mu m$ beam waist, whereas 
          $\epsilon < 0.0001$ is reached when 
          the optimization allows for amplitudes up to 
          $\unit[325]{kV/m}$, or a corresponding power of 
          $4\times \unit[89]{mW}$.
        }
	\label{fig:highEnergyOptimised}
\end{figure}
\Cref{fig:highEnergyOptimised} shows the Bell state, i.e.,
entanglement error, obtained when rescaling the duration 
to $\unit[1]{ms}$, for different field strengths
$E_{r/b}$ and heating rates
(shown in red, orange, and grey, respectively). 
A fidelity of $F=99\%$ is reached for all regarded heating rates, requiring 
field strengths between $E_{r/b}
= \unit[31]{kV/m}$ ($\kappa_{h}=2\pi\times\unit[0]{\rate}$) and 
$E_{r/b}=\unit[38]{kV/m}$ ($\kappa_{h}=2\pi\times\unit[10]{\rate}$).
The next threshold, $F=99.9\%$,
is only crossed for $\kappa_h=0$ and 
$\kappa_{h} = 2\pi\times\unit[1]{\rate}$ at field strengths of 
$E_{r/b}=\unit[100]{kV/m}$ and $E_{r/b}=\unit[125]{kV/m}$, corresponding 
to total powers of $P\approx 4\times\unit[8.3]{mW}$ 
and $P\approx4\times\unit[13]{mW}$, respectively.
Obtaining this fidelity requires detunings on the order of $\unit[6]{THz}$.
The highest threshold,  $F=99.99\%$, is reached for 
$\kappa_{h}=2\pi\times\unit[0]{\rate}$ whilst requiring field strengths 
of the order of $E_{r/b} \approx \unit[325]{kV/m}$ 
corresponding to a total power of $4\times \unit[89]{mW}$.
This is about two and a half times more power than for the gate based approach
in Ref.~\cite{ozeri2007errors}.
For neglible heating, the required field strength per beam can be reduced by using three instead of four beams to drive the two sideband transitions (blue curve in Fig.~\ref{fig:highEnergyOptimised}). 
This finding illustrates that parameter optimization is prone to trapping in local optima, in particular for a larger number of optimization parameters~\cite{GoetzPRA16b}.
We attribute the improvement to the fact that omission of one beam reduces the inadvertent scattering. More specifically, the extra constraint on the beam detunings
appears to aid the optimization algorithm in finding a 
configuration for which the contribution of each beam 
towards the scattering error is distributed in a more favorable
way than in the four beam setup.

\section{Conclusions}
\label{sect:Conclusions}

We have addressed the problem of additional noise sources that limit fidelity 
and speed of  dissipative entanglement generation. 
Combining quantum optimal control theory~\cite{glaser2015training}
with the effective operator approach~\cite{reiter2012effective},
we have shown how to improve both fidelity
and speed for the example of entangling two
hyperfine qubits in a chain of trapped ions
~\cite{lin2013dissipative}. 
The detrimental noise source in this case 
is undesired spontaneous decay brought
about by the sideband laser beams that
are necessary for coupling the qubits
~\cite{lin2013dissipative}. 
This decay leads to the irrevocable 
loss of population from the hyperfine 
subspace of interest.
Whilst the undesired spontaneous decay 
cannot be eliminated entirely,
an optimal choice of the experimental parameters increases the
fidelity from 76\% to 85\%, with minimal changes to the
setup. Key to the improvement is  optimization of the sideband laser
beam polarizations which enter the decay rates of each individual
hyperfine level. Due to their interdependence,  the various parameters
of the experiment need to be retuned when changing the
polarization. 
The two-stage optimization process that we have
developed here can easily resolve this issue, demonstrating the power
of numerical quantum control.  

Further limitations to fidelity and speed can be identified graphically,
by visualizing the connections between states
due to the various field-driven transitions. 
This allows to  qualitatively trace the flow of population and shows that, 
depending on the
relative transition rates, population can get trapped in states other than 
the target state or be transferred out of the target state by an unfavourable 
transition. The latter in particular implies that the target state does not 
fully coincide with the steady state of the evolution. 
In order to overcome this limitation, we suggest to adapt the entanglement scheme
presented in Ref.~\cite{bentley2014detection} to using two sideband transitions. 
Of course, adding a second sideband makes the
suppression of the error due to detrimental spontaneous emission even more 
important. Our optimization method had no difficulty to cope with this task, 
despite the increase in the number of tunable parameters. 

Analysis aided by the graph of connected states for the two-sideband scenario 
reveals that the limitations of the original scheme of
Ref.~\cite{lin2013dissipative} can indeed be overcome, whilst
providing additional advantageous properties such as higher
entanglement speed and  
inherent cooling. This offers the possibility of reducing the
complexity of the experiment by removing the need for sympathetic
cooling and all sources of error that come with it. The entangled, or, 
Bell state fidelity that we predict for  
the two-sideband scenario strongly depends on the heating rate. It can
be as high as $98\%$ under conditions similar to those of
Ref.~\cite{lin2013dissipative}, in particular in terms of the
available laser field strengths. The maximum attainable fidelity is
primarily limited by the heating rate of the motional mode that is
used to couple the qubits. It dictates the timescale at which the
sideband transitions must take place. Weaker sideband transition rates
in turn enable better suppression of spontaneous emission errors. 
Whilst a fidelity of $98\%$ represents an order of magnitude
improvement over the originally obtained fidelity, execution of most
quantum protocols requires fidelities in excess of $99\%$. These
could be achieved through experimental refinements, such as ion traps
with weaker anomalous heating, 
more powerful sideband lasers~\cite{ozeri2007errors}, or use of
optical instead of hyperfine
qubits~\cite{schindler2013quantum,barreiro2011open}. 

Provided that heating rates can be made negligible,
for instance by implementing the two-sideband
scheme on the stretch rather than the center of mass mode of the two ions,
one may wonder whether
spontaneous emission ultimately limits the performance of dissipative
Bell state preparation. Spontaneous emission can be reduced by using larger
detunings which in turn requires more laser power or longer durations. Compared to
gate-based entanglement preparation~\cite{ozeri2007errors}, we find,
for the same laser power of $4\times 36\,$mW into 20$\,\mu$m beam
waist as in Ref.~\cite{ozeri2007errors}, the entangling duration to
realize  a Bell 
state fidelity of 99.99\% to be increased from 10$\,\mu$s to 4.6$\,$ms
in an ideal trap. The advantages of the dissipative approach, in
particular its inherent robustness against noise, might easily outweigh this
time requirement, making dissipative entanglement production a viable
resource for quantum information protocols. Consider, for example,
carrying out primitives such as  gate teleportation. This 
could be driven by an entanglement machine that produces 200 pairs/s in serial
mode (per node) or  output one pair per  10$\,\mu$s when run with 250
nodes in parallel. A further speed up is possible by using more 
laser power.

Our study provides a first example for how to use quantum optimal
control theory to push driven-dissipative protocols to their ultimate
performance limit, despite imperfections in a practical
setting. Performance limits include, in addition to maximal fidelity,
also the highest speed. Here, we have obtained a speed up of about a
factor of four compared to Ref.~\cite{lin2013dissipative}. Speed is of particular concern when scaling up
entanglement generation since some undesired decoherence rates are
known to scale with system size~\cite{monz2011fourteen}. Deriving the
fastest possible protocol is therefore key if dissipative generation
of many-body entanglement~\cite{reiter2016scalable} is to succeed. As
we have shown, optimal control theory is a tool ideally suited to tackle this
task, and targeting a multipartite entangled state is a natural next step. 

The optimal control theory framework for dissipative entanglement generation that we have introduced here is not limited to the specific example of trapped ions. In fact, our technique is applicable to generic multi-level quantum systems in the presence of dissipation for which the time evolution can be obtained within reasonable computation time~\footnote{Typically of the order of one hundred propagations are necessary to converge the optimization}. This includes also systems with multiple steady states~\cite{KrausPRA08,AlbertPhD} which would be interesting for e.g. quantum error correction, or systems with  
non-Markovian dynamics such as solid state devices~\cite{Doucet18}. 
In the latter case, the  generalization requires the combination of the present optimization approach with one of the methods for obtaining non-Markovian dynamics~\cite{deVegaRMP17}, such as partitioning the environment into strongly and weakly coupled parts~\cite{ReichSciRep15}. Non-Markovianity has been shown to assist entanglement generation in coupled dimers subject to dephasing noise~\cite{HuelgaPRL12}. Our approach would allow to investigate, for more complex systems  and other types of dissipation,
whether non-Markovianity is beneficial or detrimental to the speed and overall success of entanglement generation.

\begin{acknowledgments}
  We thank Dave Wineland for discussions and Yong Wan and Stephen 
  Erickson for helpful suggestions on the manuscript. 
  Financial support
  from the Deutsche Forschungsgemeinschaft (Grant No. KO 2301/11-1) 
  is gratefully acknowledged.
  Florentin Reiter acknowledges support by a Feodor-Lynen fellowship
  from the Humboldt Foundation.
\end{acknowledgments}

\end{document}